%% file: usenix_template/usenix2019_v3.tex
\documentclass[letterpaper,twocolumn,10pt]{article}
\usepackage{usenix2019_v3}

\usepackage{tikz}
\usepackage{amsmath}

\usepackage{caption}
\usepackage{subcaption}

\usepackage{multirow}

\usepackage[table,xcdraw]{xcolor}

\usepackage{pifont}

\begin{document}

\input{body}

\bibliographystyle{plain}
\bibliography{reference}

\clearpage
\appendix
\input{pages/appendix}
\end{document}

%% file: body.tex
\input{macros}

\input{pages/title}
\input{pages/abstract}
\input{pages/intro}

\input{pages/goals}
\input{pages/design/design}

\input{pages/implementationOnDNS/implementationOnDNS}
\input{pages/locBasedServices/locBasedServices}
\input{pages/implementation}
\input{pages/evaluation/evaluation}

\input{pages/relatedWork}
\input{pages/conclusion}
\clearpage

%% file: macros.tex
\newcommand{\systemname}{OpenFLAME}

\newcommand{\website}{\url{https://anon-vps.github.io/}}

\newcommand*\circled[3]{\tikz[baseline=(char.base)]{
            \node[shape=circle,fill=#1,inner sep=#3] (char) {\textcolor{white}{#2}};}}
\def\checkmark{\tikz\fill[scale=0.4](0,.35) -- (.25,0) -- (1,.7) -- (.25,.15) -- cycle;}
\newcommand{\cmark}{\ding{51}}
\newcommand{\xmark}{\ding{55}}

\newcommand{\reqDecentralization}{\circled{black}{I}{1pt}~}
\newcommand{\reqGlobalCoord}{\circled{black}{II}{1pt}~}
\newcommand{\reqOwnership}{\circled{black}{III}{1pt}~}

\newcommand{\hostGeodomains}{\circled{caribbeangreen}{A}{1pt}}
\newcommand{\hostDNSreg}{\circled{caribbeangreen}{B}{1pt}}
\newcommand{\hostScan}{\circled{caribbeangreen}{C}{1pt}}
\newcommand{\hostWaypts}{\circled{caribbeangreen}{D}{1pt}}

\definecolor{bleudefrance}{rgb}{0.19, 0.55, 0.91}
\definecolor{caribbeangreen}{rgb}{0.0, 0.8, 0.6}


\newcommand{\sagar}[1]{}
\newcommand{\notes}[1]{}
\newcommand{\srini}[1]{}
\newcommand{\agr}[1]{}

%% file: pages/title.tex
\date{}

\title{\systemname{}: A Federated Spatial Naming Infrastructure}

\author{
{\rm Sagar Bharadwaj}\\
Carnegie Mellon University
\and
{\rm Ziyong Ma}\\
Carnegie Mellon University
\and
{\rm Ivan Liang}\\
Carnegie Mellon University
\and
{\rm Michael Farb}\\
Carnegie Mellon University
\and
{\rm Anthony Rowe}\\
Carnegie Mellon University
\and
{\rm Srinivasan Seshan}\\
Carnegie Mellon University
} 

\maketitle

%% file: pages/abstract.tex
\begin{abstract}

Spatial applications, i.e., applications that tie digital information with the physical world, have improved many of our daily activities, such as navigation and ride-sharing. This class of applications also holds significant promise of enabling new industries such as augmented reality and robotics. The development of these applications is enabled by a system that can resolve real-world locations to names, or a spatial naming system. Today, mapping platforms provided by organizations like Google and Apple serve as spatial naming systems. These maps are centralized and primarily cover outdoor spaces. We envision that future spatial applications, such as persistent world-scale augmented reality, would require detailed and precise spatial data across indoor and outdoor spaces. The scale of cartography efforts required to survey indoor spaces and their privacy needs inhibit existing centralized maps from incorporating such spaces into their platform.

In this paper, we present the design and implementation of \systemname{}\footnote{OpenFLAME stands for Open Federated Localization and Mapping Engine}, a federated spatial naming system, or in other words, a federated mapping infrastructure. It enables independent parties to manage and serve their own maps of physical regions. This unlocks scalability of map management, isolation, and privacy of maps. The discovery system that identifies maps hosted at a given location is a primary component of our system. We implement \systemname{} on top of the existing Domain Name System (DNS), which enables us to leverage its existing infrastructure. We implement map services such as address-to-location mapping, routing, and localization on top of our federated mapping infrastructure. 
\end{abstract}

%% file: pages/intro.tex
\section{Introduction}
\label{sec:intro}

Spatial applications, i.e. applications that utilize spatial data and tie digital information to the physical world, have revolutionized many industries such as navigation, ride-sharing, product delivery, and transportation. They hold the promise of enabling emerging industries such as world-scale augmented reality and autonomous robots. While the vision of spatial applications is promising, the reality is that such applications are hard to build today because they lack an underlying infrastructure to discover and reference spatial content. 

A well-designed naming infrastructure---a system for identifying and discovering entities in the system---would enable spatial applications to easily reference relevant content. The DNS, for example, was a key enabler of the Web and the Internet at large. It provided a simple mechanism for converting human-readable names (domain names) to server IP addresses. Similarly, we believe spatial applications need a spatial naming system to relate human-readable names (e.g., Louvre Museum) with real-world locations (e.g., $48^0 N$, $2.3^0 E$) and the content associated with those locations (e.g., museum collections). Since this naming system translates names to physical locations, we use the term \textit{map} to refer to it. 

The structure of a naming system plays a critical role in shaping and limiting the functionality of any distributed system built upon it. For example, the way a naming system handles the addition of new entities can introduce bottlenecks that hinder scalability and maintenance. The expansion of the Web and the broader Internet in their early days was partly due to the federated and pseudo-decentralized nature of the DNS, which enabled organizations to independently manage their participation on the network. This relationship between naming systems and application constraints also applies to spatial applications.
For example, the centralized and single-owner nature of today's spatial naming systems or digital maps, such as Google and Apple maps, limits their functionality. 

An example of the limitations that centralized maps face is that only information gathered and exposed by organizations maintaining them is available to applications. Extending spatial applications indoors is a use case that highlights the importance of a federated mapping infrastructure. Indoor maps contain sensitive information that needs to be owned and controlled by the owner of the physical space. For example, many organizations would benefit from providing accurate indoor map of their private offices and integrating it with outdoor maps for applications such as office navigation for their employees, but may not be willing to publicly host detailed maps. In addition, the effort required for the storage and cartography of indoor spaces far exceeds that of outdoor maps~\cite{commercialBuildFactsheet} and surveying this space will likely be impractical for any single centralized organization. While crowd-sourcing into centralized maps is a possible solution, many organizations would not cede control of their maps to centralized organizations.

In this paper, we describe the design and implementation of \systemname{}, a federated mapping system. The fundamental units of \systemname{} are \textit{map servers}---independent systems deployed by potentially disparate parties that provide map data and services confined to a physical region. 
\systemname{} needs to provide the means to efficiently discover map servers relevant for a region and combine information from these servers to support services like name-to-location translation, routing, and localization that are essential to spatial applications. Our contributions are:



\begin{itemize}
    \item We present the first design of a federated mapping infrastructure. Our infrastructure can support heterogeneous maps, has a low barrier to entry, ensures map privacy, and enables fine-grained access control (\S~\ref{sec:design}).
    
    \item We present a DNS-based implementation of the federated mapping system (\S~\ref{sec:implementationOnDNS}). The advantage of leveraging the DNS is that we have access to its existing deployment and caching mechanisms. We describe how we convert location-based queries to DNS lookups. We also show how arbitrary regions representing map boundaries can be registered on the existing DNS as name records. 
    
    \item We implement location-based services on top of our mapping infrastructure (\S~\ref{sec:buildingLocServices}). Specifically, we describe our implementations for address-to-location mapping (or geocoding)~\ref{subsec:geocodingExplanation}, routing~\ref{subsec:routingService}, and localization~\ref{sec:localization}.
    
\end{itemize}

We evaluate the discovery system and map services and compare them to centralized services in \S~\ref{sec:evaluation}.%
\footnote{We demo some tools built for \systemname{} on \website{}. We highly encourage reviewers to try them out.}

%% file: pages/goals.tex
\section{Example application}
\label{sec:exampleApp}

To better understand the needs of future spatial applications, we start by describing what we consider a typical such application, campus navigation. While we use this application to describe our design later, our goal is to support all future spatial applications. 

\begin{figure}
    \centering
    \includegraphics[width=0.95\columnwidth]{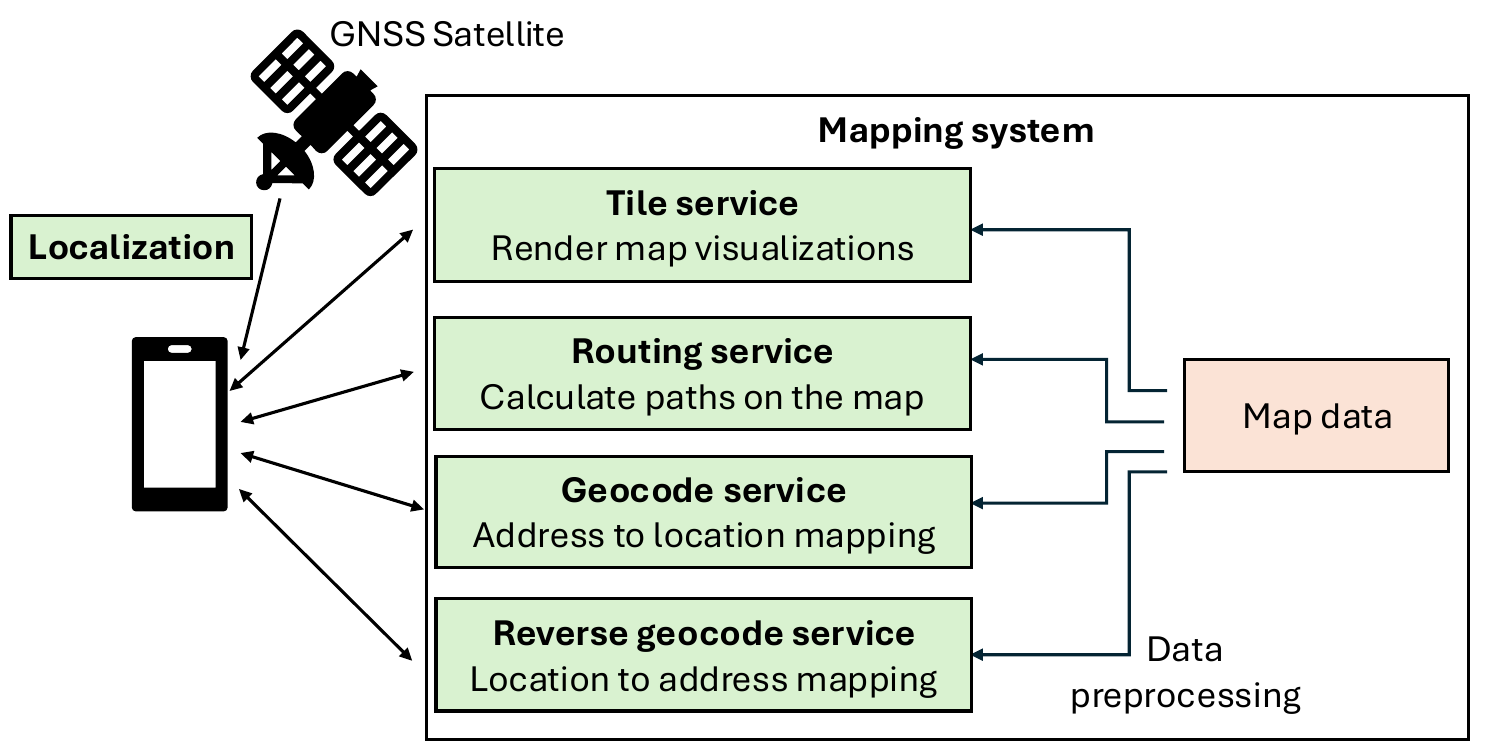}
    \caption{Centralized map architecture.}
    \label{fig:centralizedArchitecture}
\end{figure}

Let us consider a scenario where a user wishes to get pedestrian navigation guidance from their location in a typical city neighborhood to a specific professor's office at a nearby university campus. Today, such an application would rely on a centralized mapping platform such as Google Maps to search and locate the destination. It would also use a combination of technologies to determine the location of a user, including GPS, image data from Google Street View, and WiFi/Cellular signal strength.   Figure~\ref{fig:centralizedArchitecture} shows the architecture of a centralized mapping infrastructure. It includes a centrally maintained map database which is preprocessed into different forms to enable location-based services. For example, it is processed into a graph so the routing service can run shortest path algorithms on it. Each service uses the preprocessed data to serve application requests. Such a centralized map would likely only have public landmarks and outdoor walkways. The professor's office, or the university hallways that lead to the office might not be part of the map unless the centralized service had surveyed the university. 

Ideally, we would like the application to provide precise visual guidance along all steps of the path. Unfortunately, existing applications fail to meet this objective in multiple ways -- failing to provide precise guidance when localization is inaccurate, or in this case, failing to provide complete guidance as the required destination is not in the map database. 

%% file: pages/design/design.tex
\section{Design}
\label{sec:design} 

\begin{figure}
    \centering
    \includegraphics[width=0.9\linewidth]{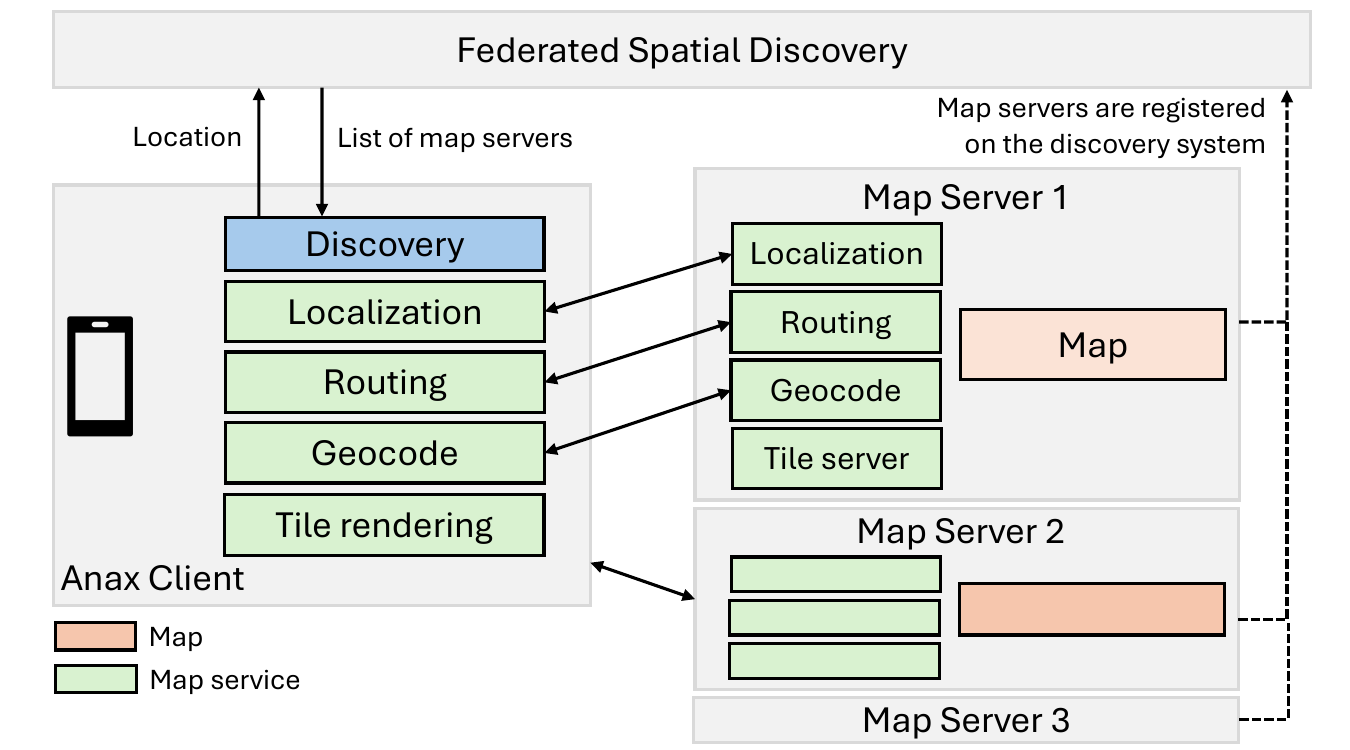}
    \caption{\systemname{} architecture.}
    \label{fig:ourArchitecture}
\end{figure}

Figure~\ref{fig:ourArchitecture} shows the simplified architecture of \systemname{}. Maps of different regions are stored on separate \textit{map servers} maintained by independent organizations. Map servers also provide location-based services on top of the maps that they store. An \systemname{} client discovers potentially zero or more map servers covering the region of interest using a federated spatial database. It then contacts these discovered servers to obtain services such as routing and localization, combining results from multiple maps when needed. Alternative design choices where the discovery system discovers spatial content or applications associated with a location are discussed in Appendix~\ref{subsec:designSpaceExploration}. In this section, we discuss our design decisions pertaining to the organization of map data (\S~\ref{subsec:dataModelAbstraction}), discovery query model (\S~\ref{subsec:discoveryQueryModel}), and security and privacy models (\S~\ref{subsec:securityAndPrivacy}). We will describe the implementation of these models in the next section (\S~\ref{sec:implementationOnDNS}).

\subsection{Data Model}
\label{subsec:dataModelAbstraction}

\subsubsection{Abstractions: Maps and Map Servers}
\label{subsec:mapServerAbstraction}


\paragraph{Map server.} A map server is a system that stores the \emph{map} of a region and provides \emph{map services} on it. For example, a university building's map server would store all the offices and navigable hallways in the building and might provide an image-based localization service to support AR applications. A map server can impose fine-grained security and privacy policies on users and applications.

\paragraph{Map.} A Map is a representation encoding relationships and attributes of spatial entities in a geographic region. While traditionally, a map refers to the visual representation of geographic features, in our context it is the data that underlies such visual representations. We do not restrict the format in which map data is stored in individual map servers.

\paragraph{Map services.} Location-based services built on top of maps are called map services.  The green boxes in Figure~\ref{fig:ourArchitecture} show some examples of map services. The tile service, for example, returns a visual representation of the map. Clients access map data only through map services.

\paragraph{Map zone.} A set of map servers are grouped to form a map zone for organizational convenience and ease of delegation. For example, the map servers of the different departments of a university form the map zone for the university.

Each map server is registered under a zone. Both map zones and servers define their own coverage---the spatial extent they are responsible for. The coverage of a map server must lie entirely within the coverage of its zone. However, the coverage of a zone may extend beyond the combined coverage of its servers, meaning that zones can include “empty” regions that are not served by any map server. A zone can delegate the responsibility of parts of its coverage down to sub-zones. A zone can be registered with one or more parent zones. 

For example, consider a university setting up its map on \systemname{}. The university first defines the coverage of its zone, which spans all buildings on its campus. This allows the university to manage its zone independently of other map zones. It registers its zone with a parent zone, e.g., the city zone. Initially, much of the university zone’s coverage may consist of empty regions not served by any map server. Over time, individual university departments can populate the zone by registering their own servers, each covering their respective areas. Within the zone, these map servers operate independently of one another and their coverages can overlap with each other. The university facilities department, for example, could set up its own sub-zone and maintain all of its maps (e.g., electricity and plumbing plans) within this sub-zone.

The fundamental unit of discovery for spatial applications is the map server. Map zones exist primarily for organizational and administrative convenience of delegation. Applications do not need to interact with zones directly. However, zones can serve as an optional filtering mechanism; i.e., applications may restrict discovery to specific zones if they wish.

\subsubsection{Organization of Map Servers and Zones}

\begin{figure}
    \centering
    \includegraphics[width=0.95\linewidth]{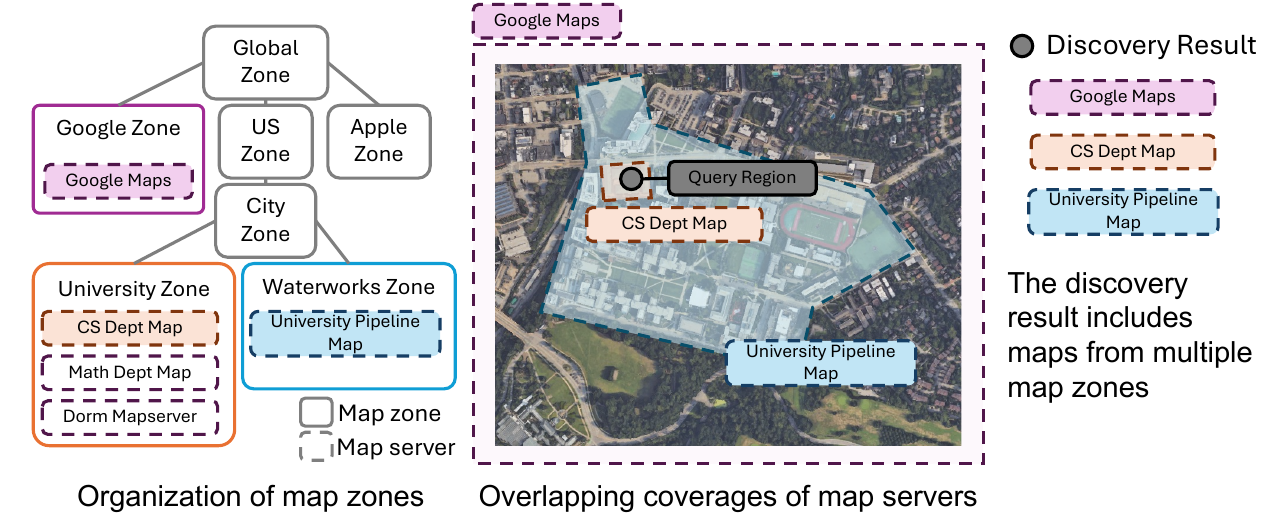}
    \caption{\systemname{} data organization and query model.}
    \label{fig:dataOrganization}
\end{figure}

Map zones form a nested inclusion hierarchy; a child zone must be completely contained within its parent zone. The coverage of map zones can overlap with each other. Map servers are the leaves of this hierarchy. They always have a parent zone, and their coverages may also overlap.

An obvious starting point for organizing the hierarchy of map zones would be to reuse existing geopolitical hierarchies such as countries, states, and cities~\cite{gibb2023earth}. However, such traditional hierarchies are not suited for our purpose. First, they are fraught with disputes. National boundaries are contested, and even property lines are frequent sources of legal conflict. Anchoring a technical system to these boundaries risks inheriting political disagreements~\cite{clark2002tussle}. Second, geopolitical hierarchies generally assume exclusive ownership, where a region belongs to exactly one parent. This prevents benign overlaps, such as a university and a commercial provider both maintaining maps of the same campus. 

The key difference between our approach and traditional hierarchies is the explicit allowance of overlaps. Multiple zones may cover the same physical region, and each can delegate to its own set of servers. This flexibility enables incremental deployment, where new maps can be attached under existing zones without requiring reorganization. Figure~\ref{fig:dataOrganization} shows an example map zone hierarchy with map servers within zones.

While overlaps enable easier integration of new maps, they also introduce challenges for map server discovery. A discovery query must now search across multiple branches of the hierarchy, increasing the complexity of discovery. We elaborate on this and discuss a feasible implementation in \S~\ref{sec:implementationOnDNS}.

\paragraph{Why hierarchy?}
Instead of organizing as a hierarchy, every map server could register its coverage with a centralized service that allows overlaps. This service could leverage existing spatial databases~\cite{mongoDB,postGIS} to store the coverage of each map server simplifying the discovery process. However, this undermines the goals of federation, since new map registrations and updates to coverage would be controlled by a single entity. A peer-to-peer design would avoid centralization but struggle to scale for complex discovery queries. A hierarchical system provides structure, supports federation through delegation, and scales more naturally with growth.

\subsubsection{Expressing Coverage}
\label{subsec:expressCoverage}

Polygons are the most intuitive representation to express the coverage of map servers and zones, but operations such as intersection checks require complex spatial indexes that are costly to build and query in distributed settings. Furthermore, polygon-based queries are not amenable to distributed caching as they rely on exact spatial boundaries. Small differences in query boundaries can cause cache misses, reducing cache reuse and increasing recomputation costs.

We can instead express coverage as a collection of primitive shapes. Existing spatial indexing systems like H3~\cite{h3} and S2~\cite{s2home} represent regions using primitive shapes. H3 and S2 decompose the world into hierarchically organized hexagons and squares respectively. The advantage of using spatial indexing systems is that discovery operations can be performed directly on the indices of primitive shapes. Moreover, queries for the same region can be cached using these indices as keys. Precisely expressing an arbitrary region might require a large number of primitive shapes. However, in practice, map coverage boundaries are rarely exact. While it is practical to guarantee that the coverage lies within a boundary, it is often difficult to assert that coverage of a zone or server extends up to every point of that boundary. We can exploit fuzzy boundaries to represent regions approximately and bound the number of primitive shapes required to cover them.

\begin{figure}
    \centering
    \includegraphics[width=0.5\linewidth]{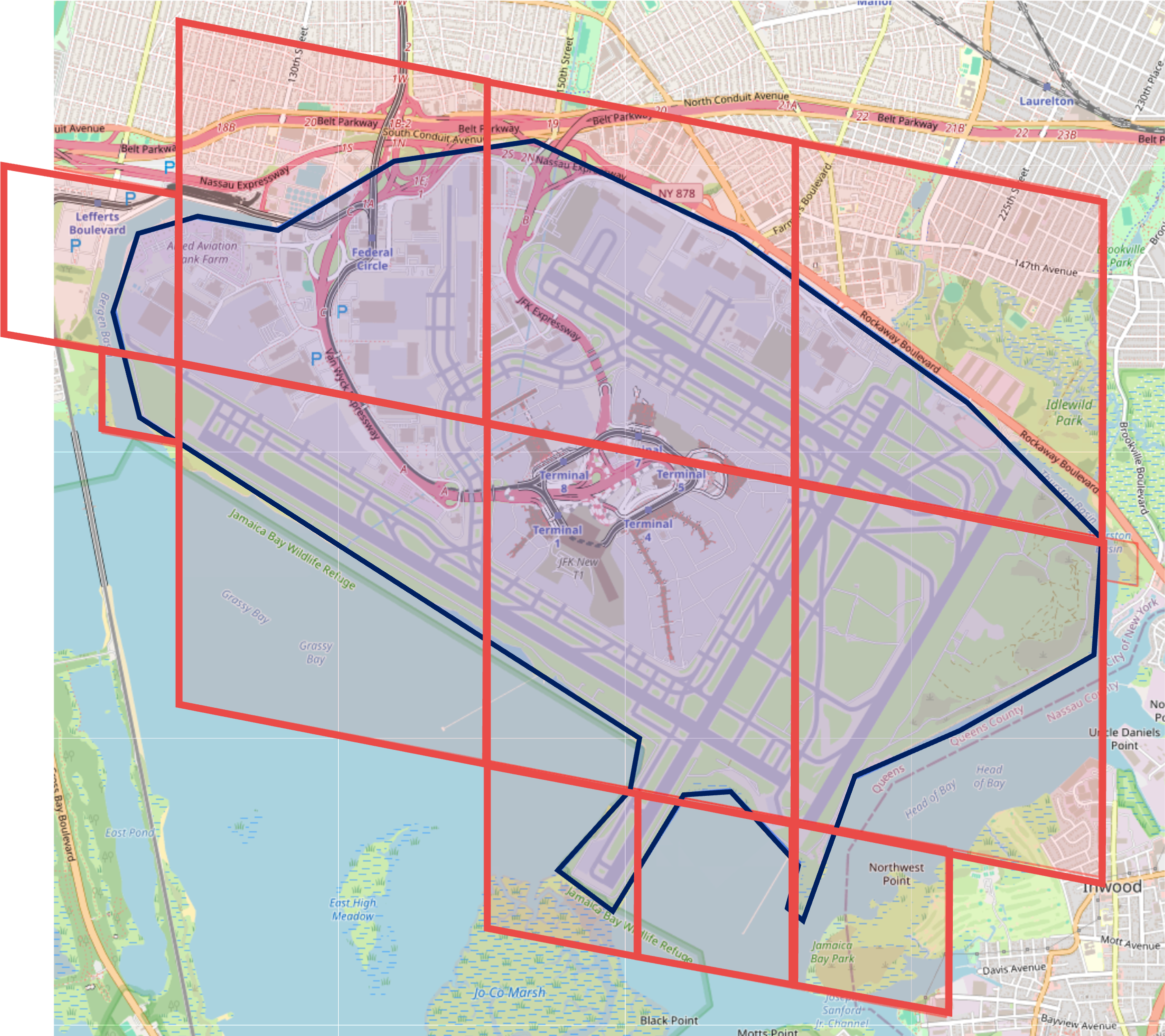}
    \caption{\textcolor{red}{S2 cell covering} for a given \textcolor{blue}{map boundary}.}
    \label{fig:s2CellsCovering}
\end{figure}

Unlike squares, hexagons do not exactly subdivide into child hexagons. As a result, the inclusion of children in parents is only approximate in H3, making S2 more suitable for representing an inclusion hierarchy of map zones. Therefore, we use the S2 spatial indexing system to express coverage as a collection of S2 cells, together with an optional altitude above sea-level. Figure~\ref{fig:s2CellsCovering} shows an example of S2 cells representing a map zone.

\subsection{Discovery Query Model}
\label{subsec:discoveryQueryModel}

A map server discovery query takes as input a list of S2 cells and optionally, altitude and a list of accepted map zones. It returns the set of map server addresses whose coverage intersects with the input S2 cells. Discovery is limited to a specific altitude, if provided. It is also restricted to map servers registered within the accepted map zones list, if provided.

\begin{equation}
\label{eqn:discoveryQuery}
\begin{aligned}
    \text{discover}&(\text{S2 Cells, [altitude], [accepted map zones]}) \rightarrow \\
    & \quad [\text{map server}_1, \text{map server}_2, \cdots]
\end{aligned}
\end{equation}

For convenience, an application can represent the search region as a 3D bounding volume, i.e., an arbitrary 2D geographic shape (e.g., a polygon or circle) together with an altitude range.
The \systemname{} client library invoked by the application converts the bounding volume to a collection of S2 cells and altitude. 

An application that already knows map zones relevant to its context can restrict the discovery process to those zones. For example, a university navigation application can restrict discovery to the university zone. If no list is provided, the discovery query is answered by first recognizing the set of map zones that cover a region followed by identifying all the map servers within these zones that cover the queried region. A consequence of allowing overlaps of map zones is that discovery queries now have to explore all branches of the map zone hierarchy that have coverage over the queried bounding region. Figure~\ref{fig:dataOrganization} shows an example where a query within a university returns maps across zones such as the city waterworks zone, university's own zone and Google Maps. An advantage of using S2 cells is that the query results can be cached across multiple layers such as client devices, Content Distribution Networks (CDNs), and Internet Service Providers (ISPs). We show in our evaluation (\S~\ref{sec:evaluation}) that, despite its complexity, the discovery query is manageable due to caching.

\paragraph{Limitations.} The discovery query model we adopt is inherently limited. It does not accept custom search terms and therefore cannot support nuanced queries such as ``find only maps of shopping complexes in a city''. Importantly, it does not support a ranking criteria to order the discovery results. As a result, querying over a region with many overlapping areas may yield large, unordered result sets. We envision that, once deployed, discovery of maps will evolve much like the Web. While the maintenance and updating of maps must be federated, much like websites on the Web are maintained independently, the discovery of maps can be centralized, analogous to a Web search engine. Centralized search engines would act as curators that crawl regions for available maps, index metadata and the services provided by map servers, and support richer search capabilities. Additionally, mechanisms such as external whitelists, blacklists, and ranking systems (e.g., crowd-sourced voting, auctions) can be employed to further tune discovery results. The discovery mechanism implementation described later in this paper (\S~\ref{sec:implementationOnDNS}) would still be essential to allow such out-of-band mechanisms to discover maps in the first place. We leave the study of nuanced map searches, filtering, and ranking mechanisms to future work. 

\subsection{Security model}
\label{subsec:securityAndPrivacy}

\subsubsection{Threat model}

Once federated mapping is deployed at scale, it will likely reveal a wide and diverse attack surface. Map servers, for example, are vulnerable to attacks such as denial of service (DoS), reflection, and amplification. We believe that existing methods on DoS protection or anomaly detection could mitigate such attacks. In this work, we choose to focus on a subset of threats unique to \systemname{} that stem from identity spoofing. Identity spoofing in this context refers to an adversary masquerading as a legitimate map server or zone, tricking applications into trusting falsified associations between regions and services. This would lead to threats such as cache poisoning and man-in-the-middle attacks. Identity spoofing may also result in denial of service for spatial applications as they could be bombarded with fake discovery results. Since caching is heavily relied upon in our model, and cached records may not always originate from the authoritative source, ensuring the authenticity of the discovery results is crucial.

\subsubsection{Chain-of-trust model for spaces}

Several existing models provide inspiration for how such identity authentication can be accomplished. For example, the Web relies on Public Key Infrastructure (PKI)~\cite{rfc5280}, where trust anchors are global Certificate Authorities (CAs) that issue digital certificates binding identities to public keys. In contrast, DNSSEC~\cite{rfc4033} and BGPSEC~\cite{rfc8205} use hierarchical chain-of-trust systems, where authority is delegated step by step (e.g., root to top-level domain to child domain), and each link in the chain validates the next.

We adopt a hierarchical chain-of-trust model for \systemname{} in which each zone signs all map servers and child zones registered with it. Clients performing discovery have configured trust anchors and validate results by ensuring that the signature chain extends to those anchors. We argue that this model is suitable for spaces as proving ownership of a map of a physical space requires proximity, which is something local or parent zones are more likely to possess than distant CAs. Furthermore, it is easier for new participants to prove their relationship to a parent zone than to undergo validation from an external CA (e.g., it is easier for a student club wanting to host a map of the university to request for validation from the university admins than from a centralized CA). Our data model is inherently hierarchical, making it natural for a parent zone that references a child to also provide a signed attestation of that child’s validity.

\paragraph{Limitations.} Hierarchical chain-of-trust is inherently easier to compromise than a tightly controlled centralized PKI, as careless or malicious child zones may issue faulty signatures compromising the rest of the chain below them. As described in \S~\ref{subsec:discoveryQueryModel}, we expect future spatial applications to rely on external curators, filter lists, and ranking mechanisms to select maps in a region. The security mechanism here serves to establish a baseline for identity.


\subsubsection{Communication with Map Servers}

Once map servers are discovered, clients can authenticate with them using standard mechanisms like password-based login or OAuth~\cite{rfc6749_oAuth2} with OpenID~\cite{openidConnect}. As map servers are discovered dynamically, it becomes increasingly important for the client to authenticate the server's identity (e.g., through X.509 certificates~\cite{rfc5280} or DNSSEC~\cite{rfc4033}). Subsequent communication between clients and servers to obtain map services can leverage whatever security protocols are most appropriate for the application context. For instance, an image-based localization service might incorporate privacy-preserving features to preserve confidentiality of both the server and the client device~\cite{speciale2019privacy, Moon_2024_CVPR, shibuya2020privacy}.

%% file: pages/implementationOnDNS/implementationOnDNS.tex
\section{Implementation on DNS}
\label{sec:implementationOnDNS}

\subsection{Why DNS?}

Existing spatial databases, such as PostGIS~\cite{postGIS} and MongoDB~\cite{mongoDB}, can support \systemname{} discovery query (i.e., Query~\ref{eqn:discoveryQuery}) off-the-shelf. They provide datatypes to represent polygons, can index geospatial coordinates, and have fast intersection operators to answer discovery queries. Hierarchy and delegation of zones can be implemented as a collection of spatial databases. However, using such databases would require developing mechanisms for federated and distributed operations, including query routing, response merging, and response caching. We believe that it is better to begin with a design that lends itself to distributed operation from the start. 

The data model of map servers and zones described in \S~\ref{sec:design} is analogous to the DNS model in several ways. First, the map zone is akin to a DNS zone representing an autonomous zone of administration. A link to a map server from a map zone is akin to a DNS record within a DNS zone. Second, DNS has a similar model of hierarchy and delegation for its zones as is described for map zones. Third, the DNSSEC model follows the chain-of-trust mechanism with pre-configured trust anchors similar to the security model described in \S~\ref{subsec:securityAndPrivacy}.

Beyond the data model, DNS infrastructure also aligns well with the needs of discovery. First, DNS has pervasive distributed caching across multiple layers such as local caches, ISP resolvers and CDNs which could benefit \systemname{} discovery queries. Second, discovery queries are exclusively read queries and do not need a full-fledged database with transaction processing. Third, map servers and zones are not expected to change very frequently so we do not need low update latencies making DNS suitable.

While the models of \systemname{} and DNS share many similarities, key differences make it non-trivial to adopt DNS directly. First, DNS does not natively support location-based queries or registrations. In \S~\ref{subsec:locToGeodomains}, we show how regions can be encoded as collections of domain names compatible with DNS. Second, unlike DNS where each zone is delegated to a single owner, \systemname{} permits overlaps between the coverages of map zones. As a result, discovery queries may need to traverse multiple delegation chains rather than a single path. We introduce additional records (\S~\ref{subsec:discovery:dnsRecords}) and query mechanism (\S~\ref{subsec:dnsQueryWorkflow}) to support such overlaps.

\input{pages/implementationOnDNS/generatingGeodomains}
\input{pages/implementationOnDNS/dnsRecords}
\input{pages/implementationOnDNS/hostingWorkflow}

\input{pages/implementationOnDNS/discoveryQuery}
\input{pages/implementationOnDNS/validation}

%% file: pages/implementationOnDNS/generatingGeodomains.tex
\subsection{Geo-domains}
\label{subsec:locToGeodomains}

\textit{Geo-domains} are the DNS-compatible addresses representing a physical region. We use the spherical geometry library S2~\cite{s2home} to generate geo-domains from a bounding region. 

S2 projects the Earth's surface onto a perfect mathematical sphere~\cite{s2overview}. The sphere is then decomposed into a hierarchy of cells called S2 Cells~\cite{s2cells}. Each S2 Cell is a quadrilateral bounded by four spherical geodesic lines---lines along the shortest path on a sphere. The highest level in the S2 hierarchy at level 0 is called a \textit{Face}. Faces divide the Earth into six large quadrilaterals. The top-level faces are recursively subdivided at each level into 4 smaller children. Cell levels range from 0 to 30, and the smallest cells at level 30 have an area of about $1 cm^2$. The six top-level cells are numbered from 0 to 5. At each level, the four child cells are numbered from 0 to 3. Each S2 cell has a 64-bit index that is essentially a concatenation of the hierarchical cell numbers in binary form with padding to extend the length to 64 bits. The S2 library also provides the region coverer algorithm---an API that returns the set of S2 cells that approximately cover a given bounding region.

\begin{figure}[t]
    \centering
    \includegraphics[width=0.35\textwidth]{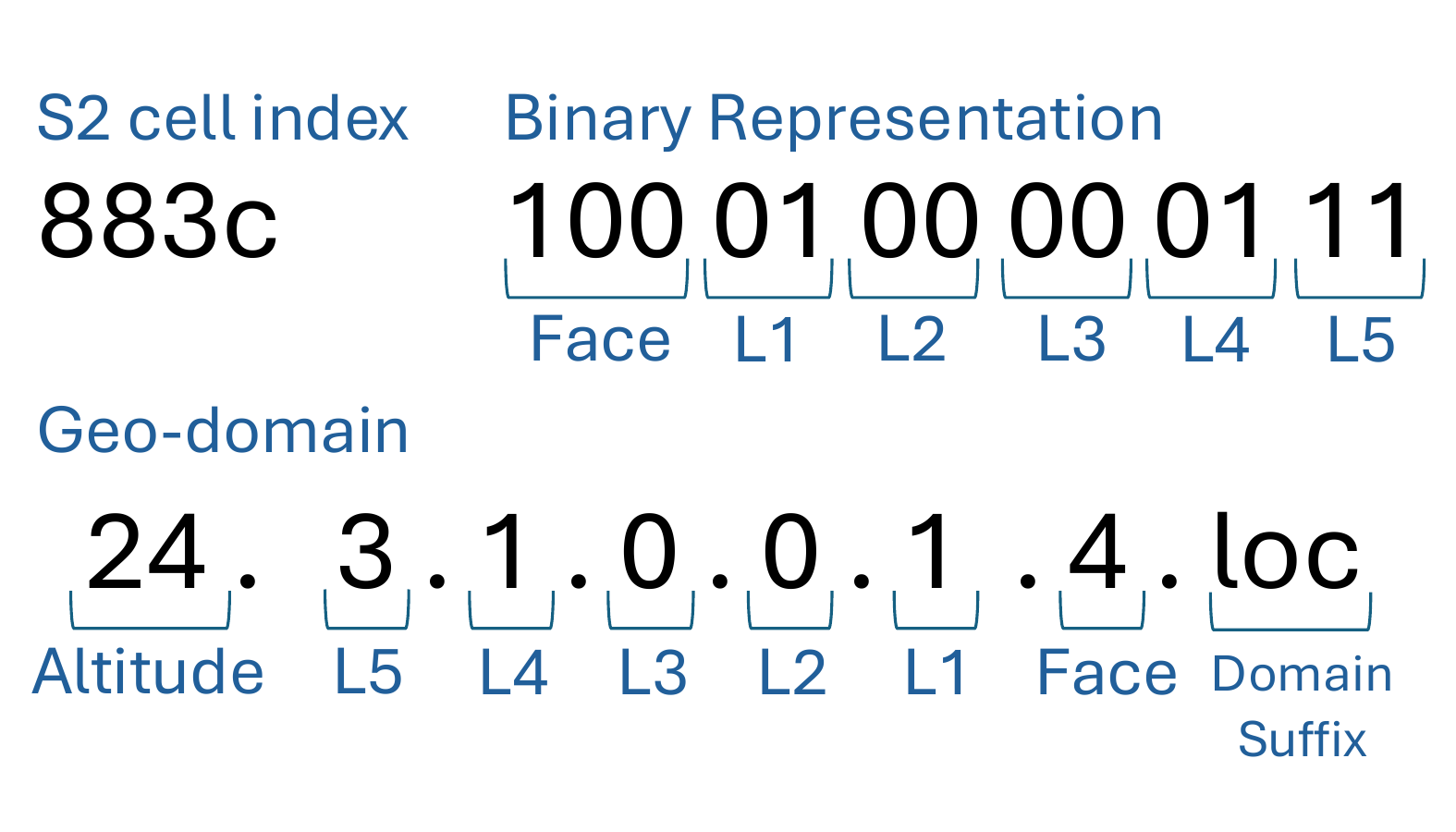}
    \caption{Converting S2 cell index to geo-domain.}
    \label{fig:discovery:s2IndexToGeodomain}
\end{figure}

The process of converting an S2 cell index to a geo-domain is shown in Figure~\ref{fig:discovery:s2IndexToGeodomain}. The S2 cell index encodes the full hierarchy of the cell up to the top-level face which we convert to a DNS-compatible format. The geo-domain is organized such that the top-level face is at the right end and the smallest cell is at the left end of the domain name. We need a \textit{domain suffix} to represent the root domain of the geo-domain. In our examples, we use \texttt{.loc}. The altitude of the S2 cell, rounded up to the nearest integer in feet, is prefixed to the geo-domain. If the altitude is unknown, the letter `$U$' is prefixed instead.

%% file: pages/implementationOnDNS/dnsRecords.tex
\subsection{DNS records}
\label{subsec:discovery:dnsRecords}

We introduce three new DNS record types to represent map servers, zones, and delegation signers. Although the structures of data in these records are same as the existing DNS records (\texttt{NS}, \texttt{CNAME}, and \texttt{DS} respectively), we introduce new records to support overlaps between zones. The owner names of all of these records are geo-domains.

\begin{itemize}
    \item \texttt{MAPSERVER} record holds the address of a map server. Its structure is the same as a \texttt{CNAME} record on traditional DNS that holds the canonical name (or alias) of a domain name. Discovery queries request \texttt{MAPSERVER} records associated with one or more geo-domains. A geo-domain can have multiple \texttt{MAPSERVER} records associated with it to allow overlaps, unlike \texttt{CNAME} records~\cite{rfc1034}.  A request for a \texttt{MAPSERVER} record informs the resolution system (e.g., DNS recursive resolver) that multiple delegation chains need to be followed to answer this request. 
    
    \item \texttt{MAPZONE} acts like an \texttt{NS} record in DNS, i.e., it delegates authority for a sub-zone to the appropriate name servers. Unlike \texttt{NS}, however, the same geo-domain may have multiple \texttt{MAPZONE} records, each leading to a different delegation chain. While DNS allows multiple \texttt{NS} records for the same domain, the underlying assumption is that they all lead to the name servers maintained by the same administrative entity with the same data~\cite{rfc1034, rfc1035}. However, for \systemname{} discovery queries, resolvers have to follow the delegation chains of all \texttt{MAPZONE} records encountered.
    \item \texttt{MAPDS} functions like a \texttt{DS} record in DNS, which DNSSEC uses to authenticate the delegation from a parent zone to a child zone~\cite{rfc4034}. The key difference is that a single geo-domain can hold multiple \texttt{MAPDS} records, each corresponding to a different delegation chain.
\end{itemize}

To ensure backwards compatibility with existing DNS servers, we wrap the above records in \texttt{TXT} records which can hold arbitrary data and is universally supported by all DNS implementations. As a result, map zones can be hosted on standard DNS nameservers without any changes. We implement map servers as web servers that implement map services such as localization, routing, and geocoding (\S~\ref{sec:buildingLocServices}). 




%% file: pages/implementationOnDNS/hostingWorkflow.tex
\subsection{Map registration workflow}
\label{subsec:hostingWorkflow}

To understand the map registration workflow, let us consider an example of hosting an airport's map on \systemname{}; for clarity, we omit signature and authenticated delegation records here and defer their discussion to §~\ref{subsec:securityDNSSECimplementation} on security.

First, a polygon is roughly drawn around the map region. S2's region-coverer algorithm is used to generate S2 cells covering the polygon. In Figure~\ref{fig:s2CellsCovering}, the blue polygon represents the map's boundary and each red square represents an S2 cell. The covering does not perfectly align with the blue polygon marking the map boundary. This is acceptable since maps on \systemname{} have fuzzy boundaries. The altitude of the map is mentioned or it is marked as unknown.

For each S2 cell, a geo-domain is generated as shown in Figure \ref{fig:discovery:s2IndexToGeodomain}. A \texttt{MAPZONE} record for each geo-domain is registered at the parent zone’s DNS nameserver (e.g., the nameserver hosting the city map zone), enabling queries within the airport to be delegated to the airport’s map zone. As described earlier, the airport can choose any parent that has coverage over the airport. These \texttt{MAPZONE} records point to the airport’s DNS nameserver. If altitude information is available, an additional set of geo-domains is registered with the altitude marked as unknown (using the prefix `U’). This allows the zone to remain discoverable even for clients that lack altitude data.

Once the airport zone is established, individual map servers can register \texttt{MAPSERVER} records in the airport's DNS server pointing to the corresponding servers providing map services. For instance, each terminal may operate its own map server. The airport zone may also delegate portions of its coverage to sub-zones by publishing \texttt{MAPZONE} records in its nameserver.


%% file: pages/implementationOnDNS/discoveryQuery.tex
\begin{figure}[t]
    \centering
    \begin{subfigure}{.2\textwidth}
        \centering
        \includegraphics[width=.95\linewidth]{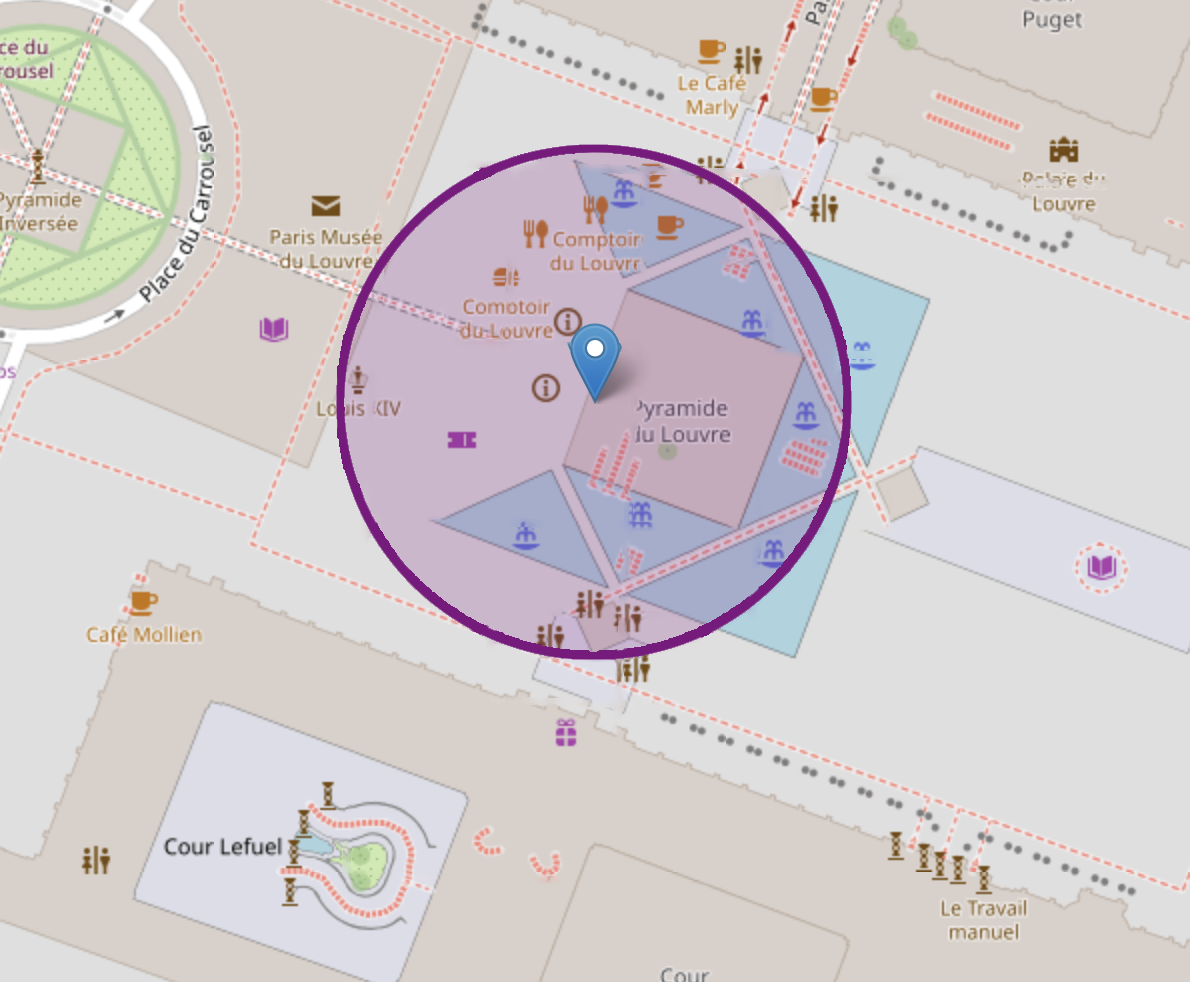}  
        \caption{Bounding region.}
        \label{fig:discovery:locationToGeodomain:circle}
    \end{subfigure}
    \hfill
    \begin{subfigure}{.2\textwidth}
        \centering
        \includegraphics[width=.95\linewidth]{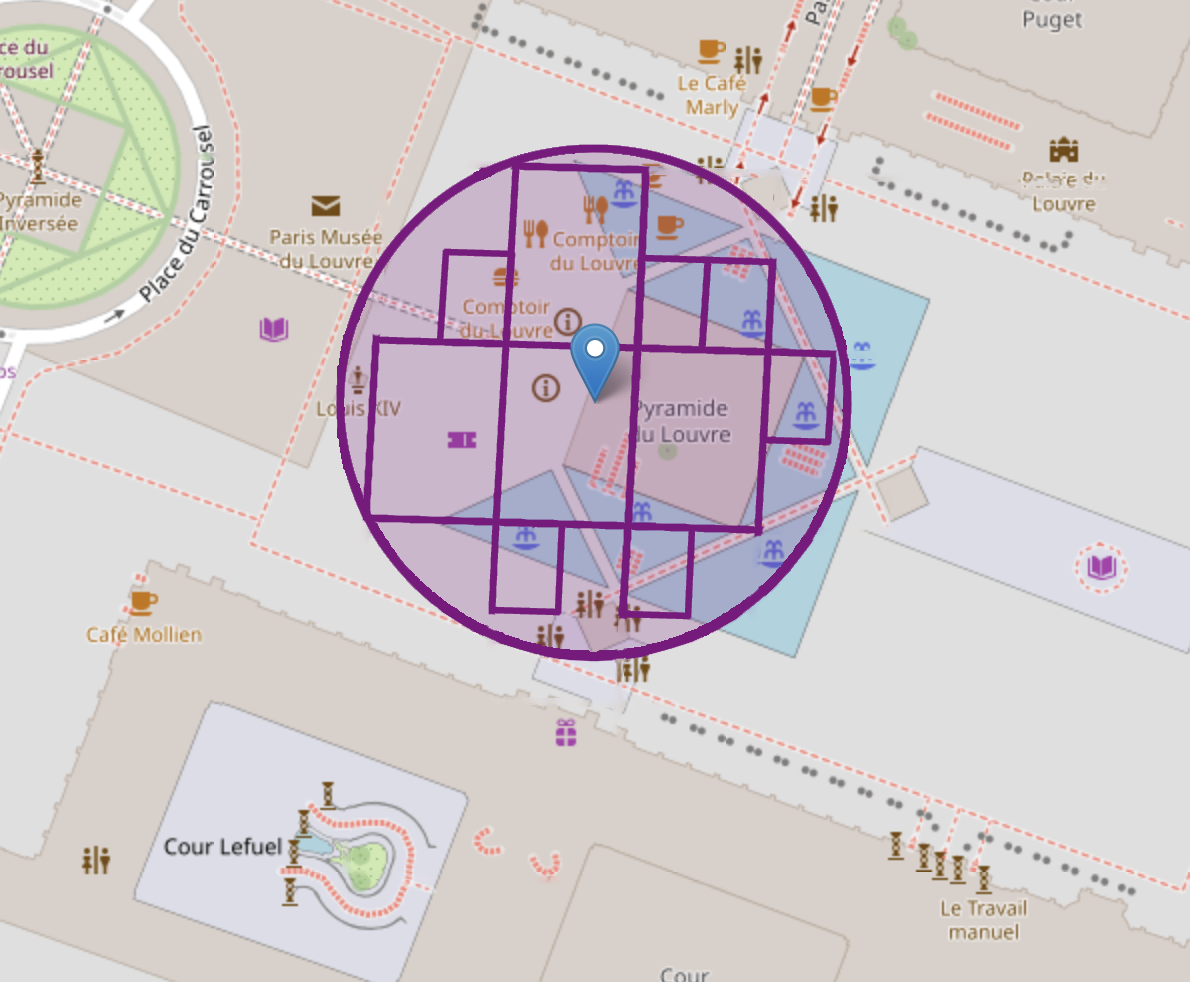}  
        \caption{Base S2 cells.}
        \label{fig:discovery:locationToGeodomain:covering}
    \end{subfigure}
    \hfill
    \begin{subfigure}{.2\textwidth}
        \centering
        \includegraphics[width=.95\linewidth]{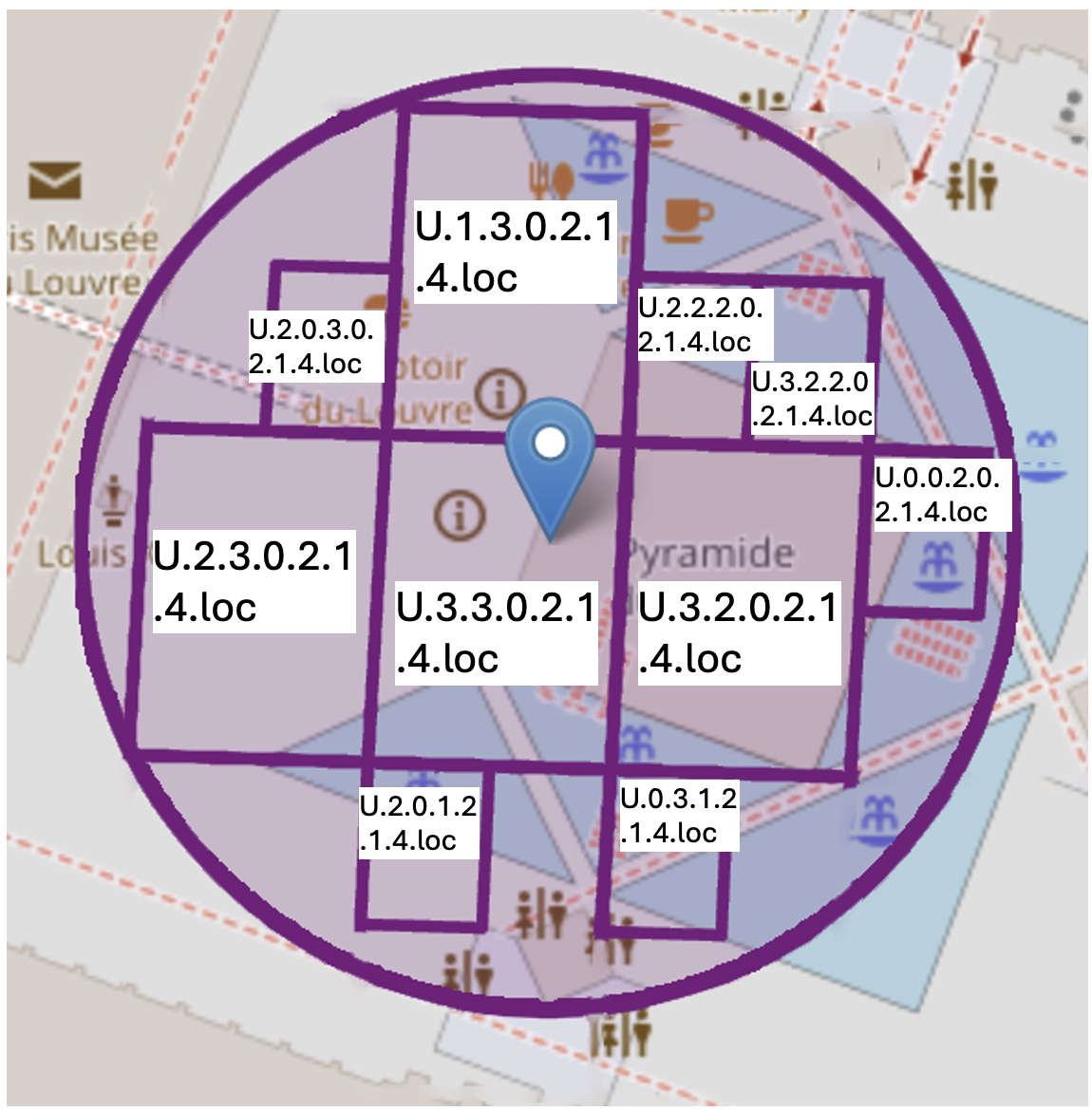}  
        \caption{Base geo-domains.}
        \label{fig:discovery:locationToGeodomain:baseDomains}
    \end{subfigure}
    \hfill
    \begin{subfigure}{.2\textwidth}
        \centering
        \includegraphics[width=.95\linewidth]{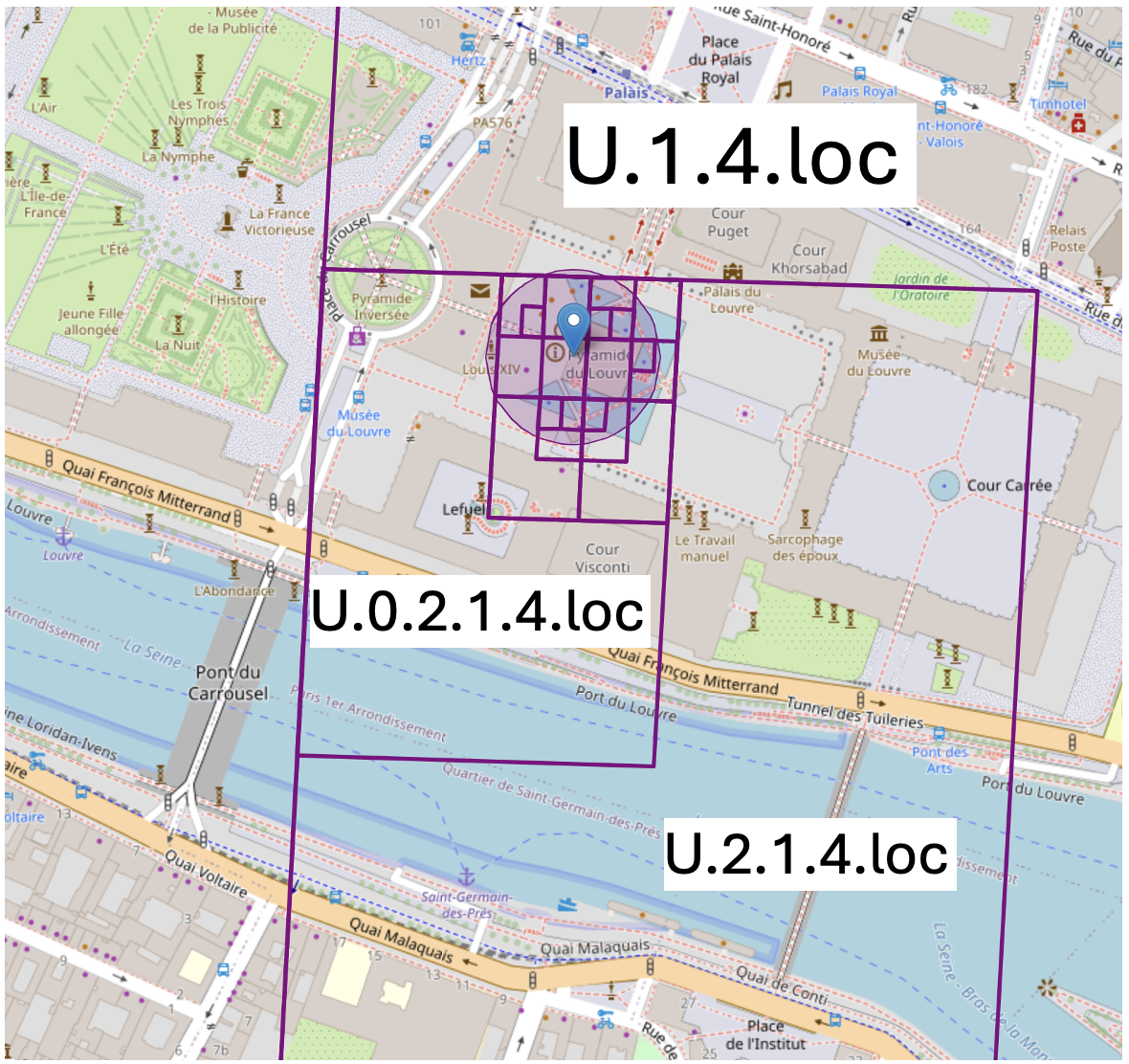}  
        \caption{Parent geo-domains.}
        \label{fig:discovery:locationToGeodomain:parentDomains}
    \end{subfigure}
    \caption{Generating geo-domains for discovery.}
    \label{fig:discovery:locationToGeodomain}
\end{figure}

\subsection{Discovery query workflow}
\label{subsec:dnsQueryWorkflow}

As described in \S~\ref{subsec:discoveryQueryModel}, an application expresses search region as a bounding volume (i.e., a 2D geographic region with altitude). The \systemname{} client first converts the request to a discovery query of the form (\ref{eqn:discoveryQuery}) and then represents it as geo-domains. Figure~\ref{fig:discovery:locationToGeodomain} shows our technique for converting a bounding region to geo-domains. 

\begin{enumerate}
    \item We use S2's interior region coverer algorithm to get a set of S2 cells that cover the 2D geographic region. We call these the \textit{base S2 cells}~(Figure~\ref{fig:discovery:locationToGeodomain:covering}).

    \item For each base S2 cell, its corresponding geo-domain is generated (Figure~\ref{fig:discovery:locationToGeodomain:baseDomains}). We call these geo-domains the \textit{base geo-domains}. 

    \item For each base geo-domain, we generate all the parent domains by removing sub-domains sequentially from the left. We retain the altitude. For example, the parents of the domain \texttt{U.1.3.5.loc} are \texttt{U.3.5.loc}, and \texttt{U.5.loc} (Figure~\ref{fig:discovery:locationToGeodomain:parentDomains}). Querying parents is essential to discover maps whose coverages are larger than the queried region. Optionally, we also add the children of base geo-domains upto a predefined level.

    \item The set of geo-domains to query includes all the base geo-domains, and their parents with duplicates removed (and optionally, their children). Optionally, a copy of the same geo-domains but with altitude marked as unknown is also queried to discover map servers not registered with a known altitude. Queries for \texttt{MAPSERVER} records of all these geo-domains are made in parallel to get a list of map servers for the queried region.
    
\end{enumerate}







\begin{figure}
    \centering
    \includegraphics[width=0.25\textwidth]{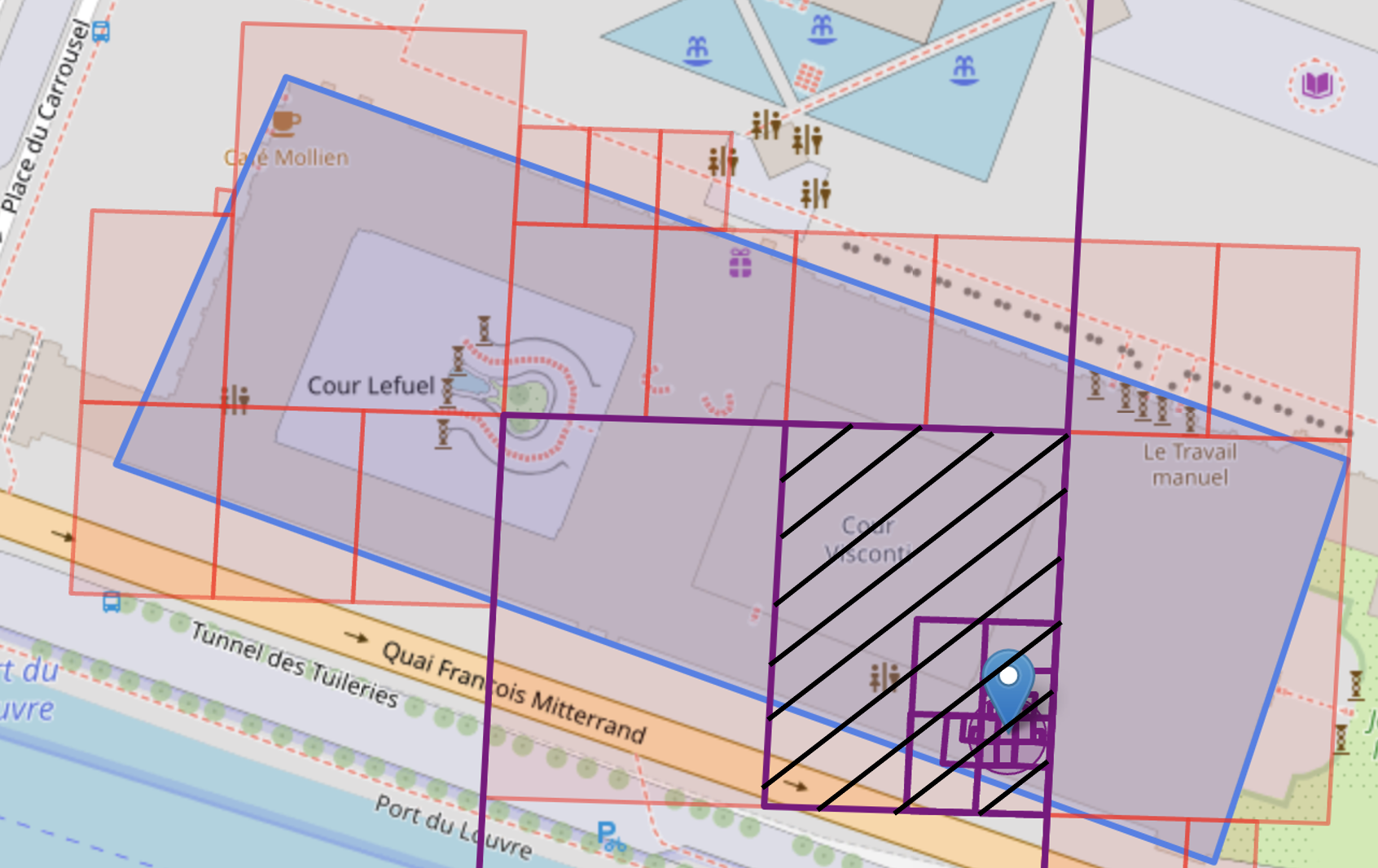}
    \caption{\textbf{Intersection} of \textbf{\textcolor{purple}{queried}} and \textbf{\textcolor{red}{registered}} geo-domains.}
    \label{fig:discovery:geodomainQuery}
\end{figure}

Figure~\ref{fig:discovery:geodomainQuery} illustrates a discovery query. Red S2 cells show the geo-domains registered for a map server serving a museum, while purple S2 cells show the geo-domains generated and queried for the input region (the small purple circle). The intersection of the registered and queried geo-domains, shown with black stripes, ensures that the address of the map server is returned by the DNS when the discovery query is made. 

\paragraph{Caching.} In every discovery cycle, the \systemname{} client typically queries about 40 geo-domains (about 10 base geo-domains and 30 parent geo-domains. See \S~\ref{sec:evaluation} for details). The ubiquitous caching mechanisms in DNS make such a large number of queries feasible. Most of the geo-domain queries are answered by the local DNS cache and never leave the device. This is because some of the top-level parent geo-domains (Figure~\ref{fig:discovery:locationToGeodomain:parentDomains}) rarely change as the queries from a client are usually limited to a small region in the real world. 


\paragraph{Fan-out queries.} In traditional DNS resolution, a query, such as one for a \texttt{CNAME} record, proceeds by following a delegation chain through \texttt{NS} records. If multiple \texttt{NS} records are present, the resolver arbitrarily selects one and follows only that chain, since all entries are assumed to lead to equivalent answers. In \systemname{}, this assumption does not hold because overlapping zones may create multiple valid delegation chains for the same geo-domain. To correctly resolve a \texttt{MAPSERVER} record, the resolver must therefore traverse all \texttt{MAPZONE} records in parallel, exploring every chain of delegation rather than selecting one. In our implementation, to maintain backwards compatibility, we leave standard DNS resolvers unmodified. Instead, the client issues \texttt{TXT} queries for the requested geo-domains. It then queries name servers in each \texttt{MAPZONE} record (wrapped in a \texttt{TXT} record) to get \texttt{MAPSERVER} records. 


%% file: pages/implementationOnDNS/validation.tex
\subsection{Security/Validation}
\label{subsec:securityDNSSECimplementation}

The security model for spaces described in \S~\ref{subsec:securityAndPrivacy} follows a chain-of-trust approach similar to DNSSEC. In DNSSEC, each record is signed using an \texttt{RRSIG} record, and these signatures are validated against public keys stored in \texttt{DNSKEY} records. Delegation between zones is authenticated using \texttt{DS} records, which establish trust from a parent to a child zone.

Our implementation reuses DNSSEC’s \texttt{DNSKEY} and \texttt{RRSIG} records as well as its standard zone-signing protocols. The key difference lies in how delegation is handled. To support multiple delegation chains for the same geo-domain, we introduce a new record type, \texttt{MAPDS}. A geo-domain may have several \texttt{MAPDS} records, each corresponding to a different delegation chain. As a result, authenticating a \texttt{MAPSERVER} or \texttt{MAPZONE} record may require traversing all relevant delegation chains.

%% file: pages/locBasedServices/locBasedServices.tex
\section{Building Map Services}
\label{sec:buildingLocServices}
 
In this section, we discuss examples of implementing map services on the \systemname{} infrastructure. We consider five key services: tile server (i.e., map visualization, \S~\ref{subsec:interactiveMapService}),  forward and reverse geocode(\S~\ref{subsec:geocodingExplanation}), routing (\S~\ref{subsec:routingService}), and localization (\S~\ref{sec:localization}). While \systemname{} can support arbitrary services, we highlight these five because they span the spectrum of inter-map interactions: tile server and reverse geocoding (location-to-address mapping) do not require interaction with other maps, forward geocoding (address-to-location search) relies on a root map to bootstrap the search, and routing and localization require combining results across multiple maps. See Appendix~\ref{appendix:heterogenousMaps} for a discussion on combining these services across maps that use different coordinate systems.

\input{pages/locBasedServices/visualization}
\input{pages/locBasedServices/reverseGeocode}
\input{pages/locBasedServices/routing}
\input{pages/locBasedServices/localization}

%% file: pages/locBasedServices/visualization.tex
\begin{figure}
    \centering
    \begin{subfigure}{0.23\textwidth}
        \centering
        \includegraphics[width=\linewidth]{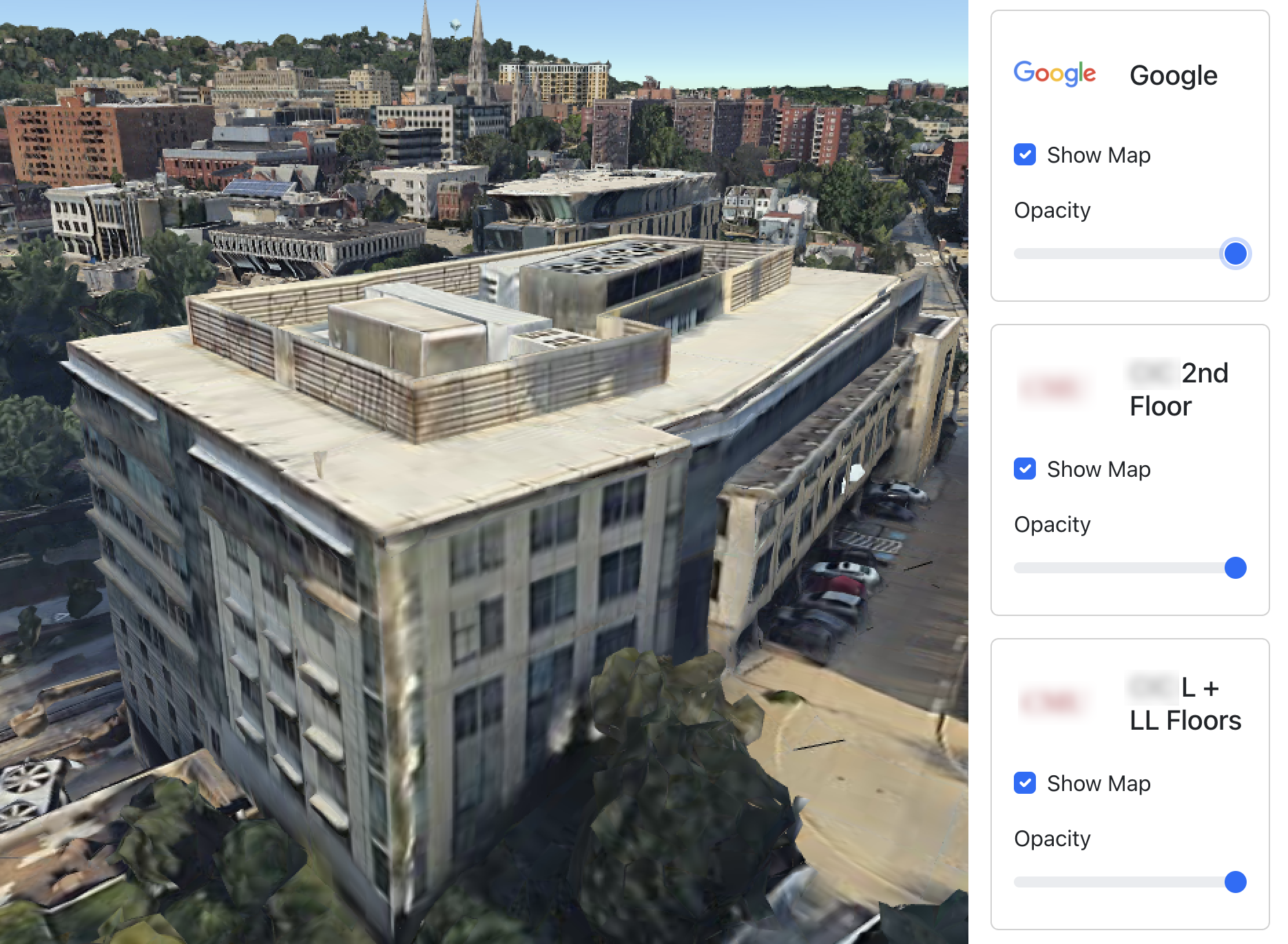}
        \caption{Outdoor view of a building from Google Maps.}
        \label{fig:outdoorMap}
    \end{subfigure}
    \begin{subfigure}{0.23\textwidth}
        \centering
        \includegraphics[width=\linewidth]{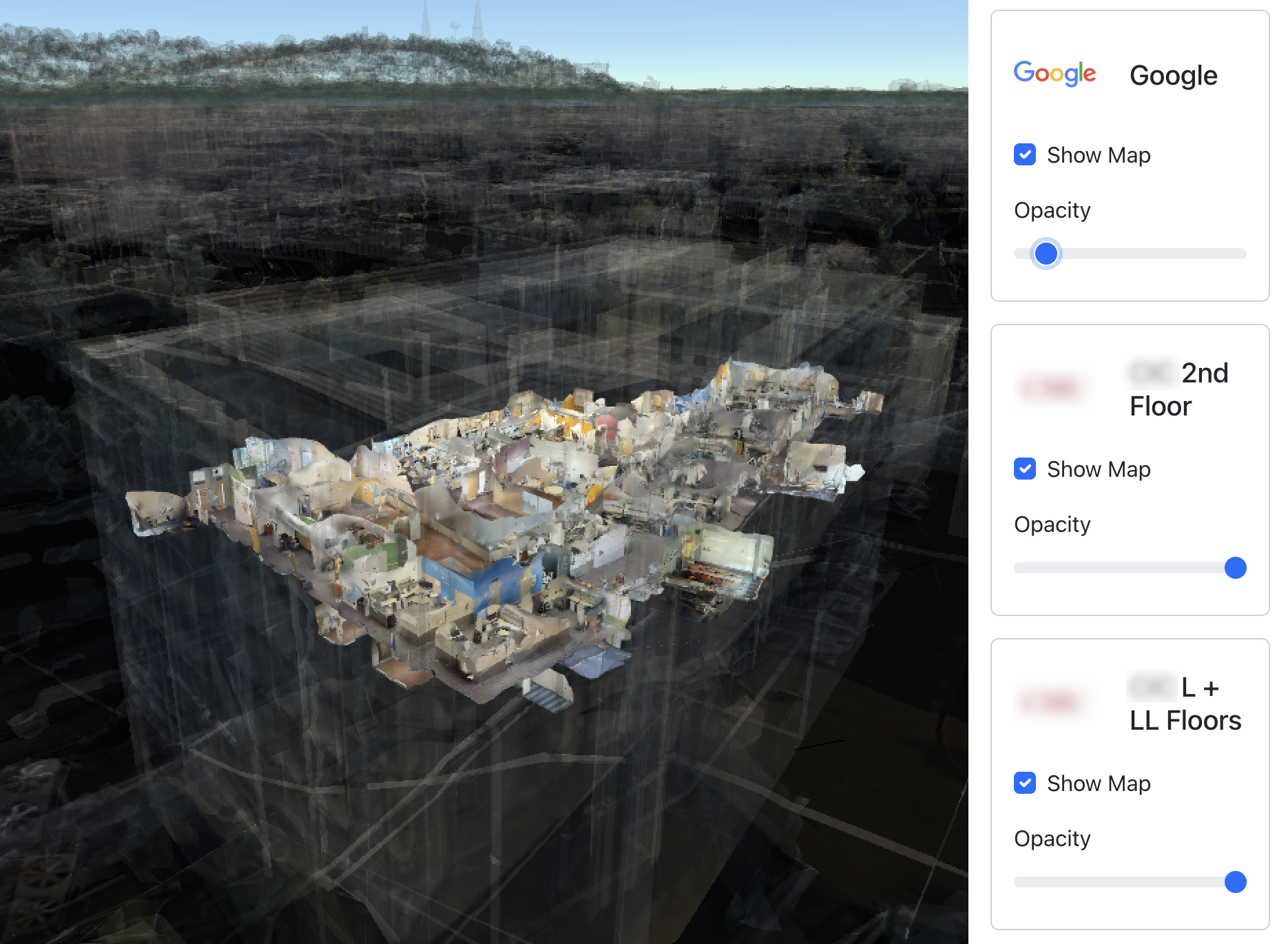}
        \caption{An indoor map of offices hosted on private servers.}
        \label{fig:indoorMap}
    \end{subfigure}
    \caption{Map visualization application built on \systemname{}.} 
    \label{fig:mapVisualization}
\end{figure}

\subsection{Tile Server}
\label{subsec:interactiveMapService}

The tile service serves the application with a visual representation of its map data. The tile service on our map servers serves 3D scans of indoors as meshes, along with a graph representing map data. We build an interactive map application that uses \systemname{} to discover map servers hosted in the user's view region, downloads tiles from them, and renders these 3D tiles alongside others for the same region. Figure~\ref{fig:mapVisualization} shows two screenshots from our application. Google Maps is registered as a server with global coverage and its 3D map tiles service~\cite{google3DTiles} provides outdoor 3D scans of buildings (Figure~\ref{fig:outdoorMap}). Figure~\ref{fig:indoorMap} shows indoor scans of offices within the building served by private university map servers that are exposed by reducing Google tiles opacity. These tiles are only visible to users with specific credentials. 

%% file: pages/locBasedServices/reverseGeocode.tex
\subsection{Forward and Reverse Geocode}
\label{subsec:geocodingExplanation}

The process of converting a text-based address to a location on the map is called forward geocoding~\cite{Goldberg2007FromTT}. In a centralized map provider, the text-based addresses of map nodes are indexed against their geolocations. So geocode involves querying this indexed database. In \systemname{}, we need a \textit{root map} with global coverage to bootstrap the geocoding process. In our case, large world-map providers such as OpenStreetMap, Google or Apple maps serve as root maps. Given a text string of a hierarchical address, the client first uses the geocode service of a root map to get the coarse location of a part of the address. The client then discovers map servers for this coarse location. It requests geocoding service from each of these map servers which search in their maps for the exact address. For example, let us consider the address ``Mona Lisa Painting, Louvre Museum, Paris, France''. While OpenStreetMap's data might not contain the location of the painting in the museum, it will give the location of the Louvre museum. \systemname{} client can then discover the Louvre map server and search for the location of the painting within that map server. 


The service that coverts a geolocation to a map node is called reverse geocode. Given a geographic location, the \systemname{} client uses the discovery system to find all the map providers in that location. Then it requests the reverse geocode service from each discovered map server and returns to the client the node that is closest to the requested geolocation. 

%% file: pages/locBasedServices/routing.tex
\subsection{Routing}
\label{subsec:routingService}

\begin{figure}
    \centering
    \includegraphics[width=0.6\linewidth]{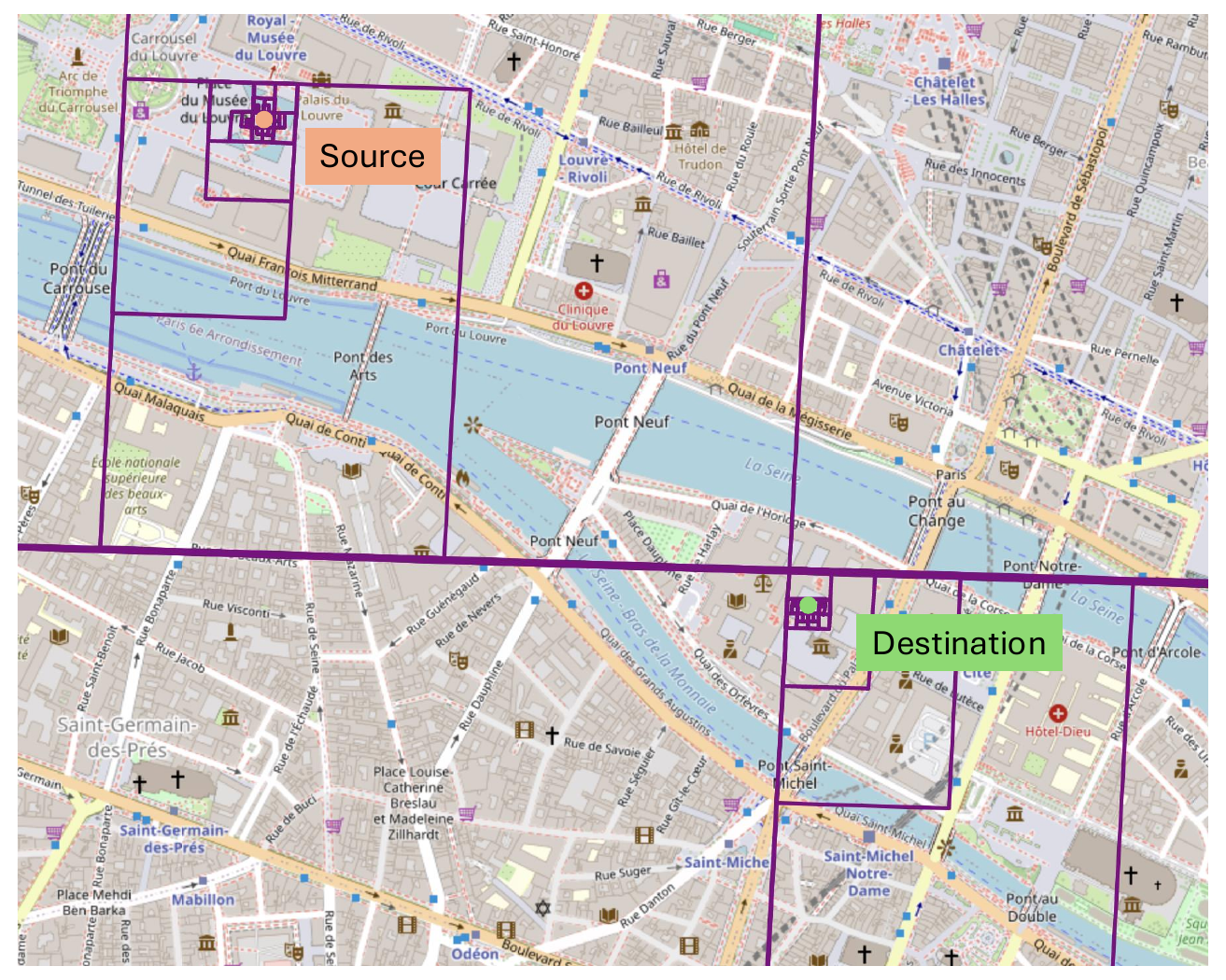}
    \caption{Geo-domains queried for routing}
    \label{fig:routingGeodomains}
\end{figure}

Routing refers to calculating a path from source to destination with optimization objectives such as minimizing travel duration, distance, or toll price. Routing in a centralized map is performed by running shortest path algorithms on the graph representation of the map~\cite{contractionHierarchies, bast2016route}. We present a simple implementation for routing on federated maps that we find is sufficient for most common cases. Given the source and destination addresses, we first use the geocode service to find their approximate geolocations. We then use the discovery system to find map providers near the source and destination. Figure~\ref{fig:routingGeodomains} shows the geo-domains queried. Querying smaller geo-domains near the source and destination locations discovers smaller and possibly more detailed maps. The client also discovers large maps that cover the region between the source and destination. The client uses a large map that spans both the source and destination to get an approximate route from the source to the destination. It then contacts smaller maps near the source and destination to refine these routes. 

Let us reconsider the campus navigation application from \S~\ref{sec:exampleApp}. The \systemname{} client would discover that Google Maps spans both the source---the street that the user is currently in, and the destination---the university premises (even though Google Maps only has coarse information at the destination). The client then requests an approximate route from Google Maps that would lead the user to the university campus entrance. It then requests the university's map server for a route from the university entrance to the professor's office. The client then stitches these two routes together to obtain a complete route from the street to the office.

This simple algorithm works for the common case where world-scale maps such Google Maps is sufficient to navigate the user outdoors, switching to other smaller maps only when indoors near the source or destination. However, this does not work for situations where the optimal path passes through smaller maps. Small detailed maps along the path are never discovered. There is a trade-off between the number of map servers contacted and the optimality of the final route. In \S~\ref{subsec:eval:routing}, we evaluate the optimality of our simple routing algorithm compared to a centralized algorithm. We leave the exploration of better routing algorithms to future work. 

%% file: pages/locBasedServices/localization.tex
\subsection{Localization}
\label{sec:localization}

Localization is the service that informs the application the position and orientation or the pose of a device with respect to a map. Centralized maps are set against the system of latitudes and longitudes so spatial applications rely on positioning systems such as GPS, WiFi, and cell towers to position the device with respect to the map. Maps in \systemname{} can be indoors (where GPS does not work reliably) and can be laid out in a separate 2D/3D coordinate system of their own. To ensure devices can localize themselves within such maps, the map servers will have to provide their own localization service. 

\begin{figure}
    \centering
    \includegraphics[width=0.30\textwidth]{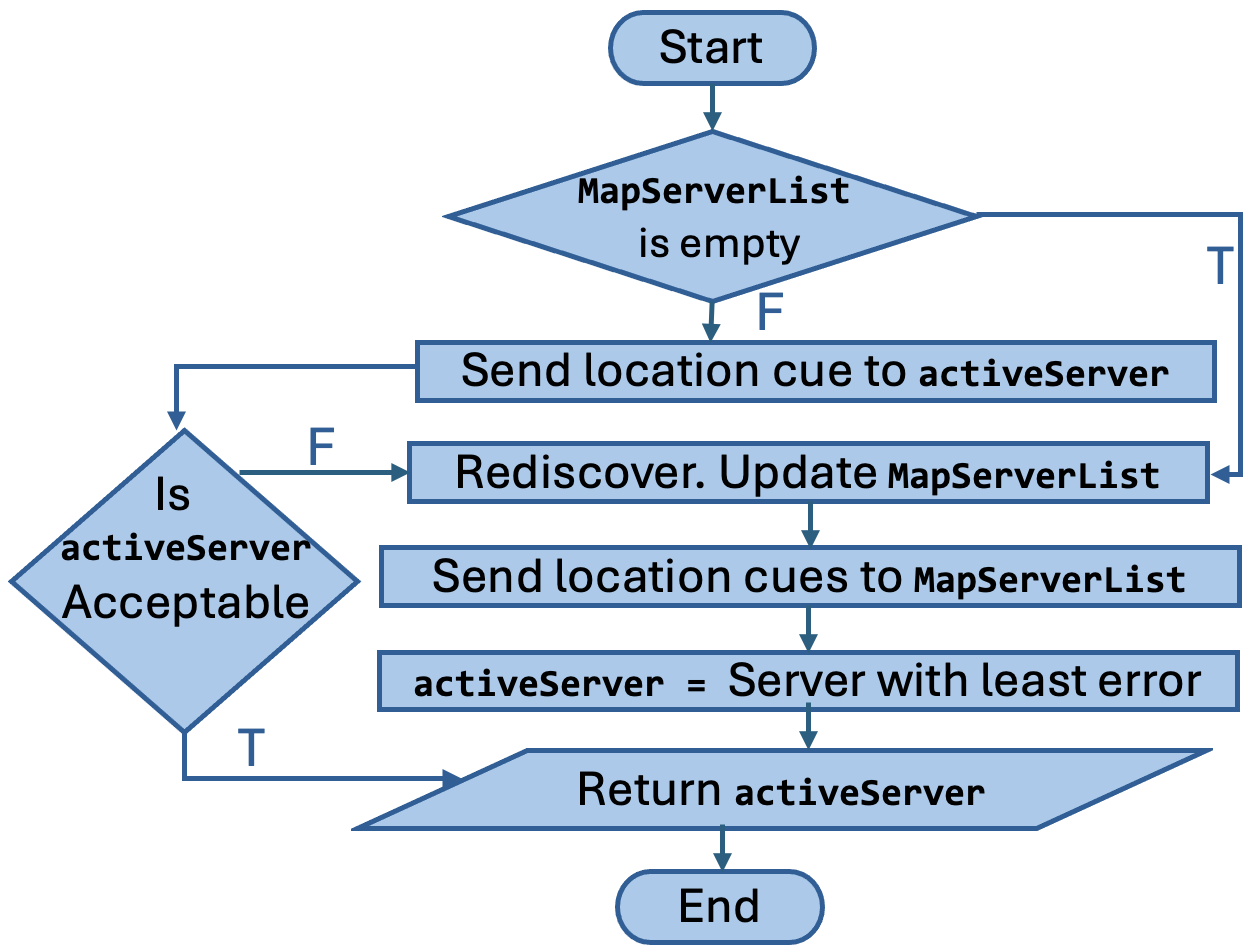}
    \caption{Localization and discovery flowchart.}
    \label{fig:localization:flowchart}
\end{figure}

Localization can be performed using various technologies including Ultra-Wideband (UWB) beacon, Bluetooth, Wi-Fi, and image-based methods. Image-based localization is a well-studied problem~\cite{klein2007parallel, mur2015orb, kendall2015posenet, mourikis2007multi}. Prior work~\cite{openflameIsmar} demonstrates image-based 6DoF localization across federated 3D maps. It uses a combination of local trajectories from device VIO sensors and remote trajectories from VPS (Visual Positioning System) servers to select the best map. It also shows how to stitch localization results across maps without exposing internal map details. Here, we implement federated localization and focus on how localization integrates with the discovery layer of \systemname{}. Figure~\ref{fig:localization:flowchart} illustrates the workflow. Generating geo-domains, making DNS queries, and sending location cues to all servers is costly in both computation and network overhead. We avoid repeating this process by exploiting temporal locality. That is, a device typically remains within the same map for some time. Thus, we can repeatedly send cues to the same server (i.e., \texttt{activeServer}) until localization quality degrades, at which point rediscovery is triggered. In \S~\ref{subsec:eval:localization}, we evaluate the client’s ability to switch between map servers.

%% file: pages/implementation.tex
\section{Implementation}

\textbf{Client library}:  We implement \systemname{} as an \texttt{npm}\cite{npm} library, bundling it for browsers to support platform-agnostic web applications. Since browsers lack direct system calls for DNS queries, we use DNS over HTTPS (DoH)\cite{rfc8484}, supported in the latest BIND9~\cite{bind9} DNS server. 

\textbf{DNS}: Our DNS server uses the BIND9 implementation. For evaluation, we run our DNS server on a machine with Intel Core i9-9820X CPU. We also provide a web-based tool to automate the process of generating geo-domains.\footnote{\website{}}


\textbf{Routing and geocoding}: Within map servers, we use Nominatim~\cite{nominatim} for geocoding and Open Source Routing Machine (OSRM)~\cite{luxen-vetter-2011} for routing.

\textbf{Interactive map}: The web application is implemented using CesiumJS~\cite{CesiumJS}, a JavaScript library for creating 3D maps. Outdoor map tiles are streamed using the Google Photorealistic 3D Tiles API~\cite{google3DTiles}. We create indoor scans using Polycam~\cite{polycam} and convert them to map tiles using Cesium Ion~\cite{CesiumIon}. These indoor map tiles are served from a Django~\cite{django} web server access-controlled using oAuth~\cite{rfc6749_oAuth2} and OIDC~\cite{openidConnect} to restrict tile service to specific logins.


%% file: pages/evaluation/evaluation.tex
\section{Evaluation}
\label{sec:evaluation}

We evaluate the DNS-based discovery system in \S~\ref{subsec:eval:discovery}. We show that most of the DNS queries in \systemname{} are cache hits and that a standard bind9 DNS deployment can support a large number of users despite discovery involving a large number of DNS queries. In \S~\ref{subsec:eval:localization}, \S~\ref{subsec:eval:routing} and \S~\ref{subsec:eval:reverseGeocode} we discuss the performance of localization, routing, and reverse geocode on federated maps and compare them with their centralized equivalents.

\input{pages/evaluation/dns}
\input{pages/evaluation/localization}
\input{pages/evaluation/routing}
\input{pages/evaluation/reverseGeocode}

%% file: pages/evaluation/dns.tex
\subsection{Discovery}
\label{subsec:eval:discovery}

\begin{figure}[t]
    \centering
    \includegraphics[width=0.7\linewidth]{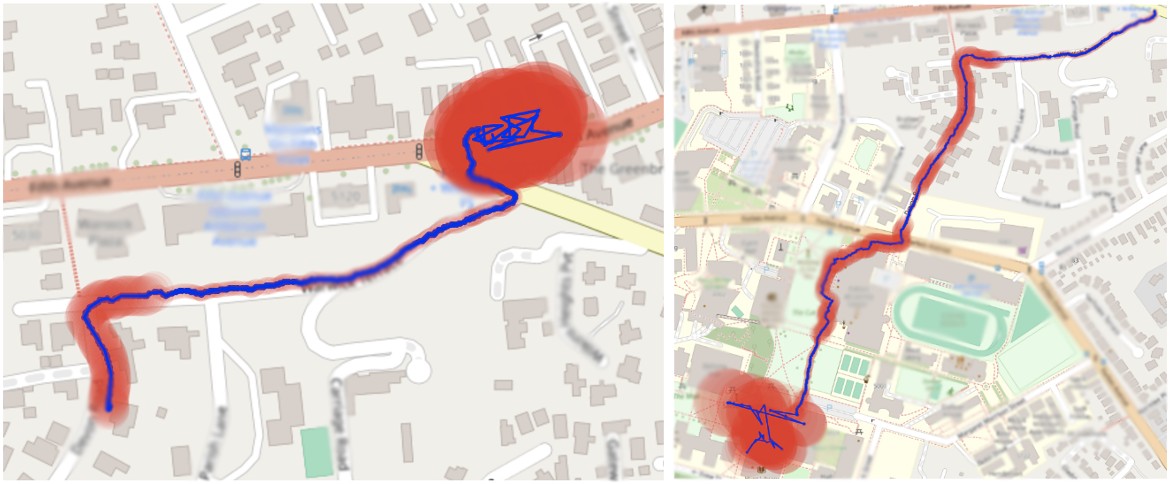}
    \caption{Examples of location traces collected.}
    \label{fig:eval:exampleLocationTraces}
\end{figure}

To evaluate the discovery phase, location traces with latitude, longitude, and 95\% confidence radius were collected by the authors. These traces are representative of the movement of a typical device that may run a location-based application on \systemname{}.  Figure~\ref{fig:eval:exampleLocationTraces} shows a few examples of the traces collected. The blue lines show the latitudes and longitudes and the red circles show the 95\% confidence radius. Note that the error is higher indoors than outdoors, as expected. We then ran the discovery phase on each of these traces.

\begin{figure*}[!htb]
    \centering
    \begin{subfigure}{.35\textwidth}
        \centering
        \includegraphics[width=0.95\linewidth]{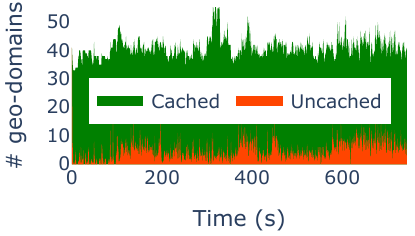}
    \caption{Number of geo-domains per discovery query over one trace.}
        \label{fig:eval:geodomainTimeseries}
    \end{subfigure}
    \hfill
    \begin{subfigure}{.25\textwidth}
        \centering
        \includegraphics[width=0.95\linewidth]{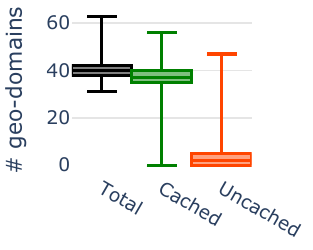}
    \caption{Geodomains queried over all traces.}
    \label{fig:eval:geodomainsBox}
    \end{subfigure}
    \hfill
    \begin{subfigure}{.25\textwidth}
        \centering
        \includegraphics[width=0.95\linewidth]{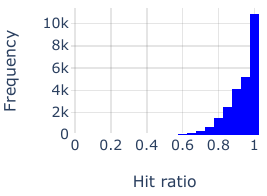}
    \caption{Distribution of cache DNS hit ratios.}
    \label{fig:eval:hitRatioDistribution}
    \end{subfigure}
    \caption{Geodomain query statistics in \systemname{}.}
    \label{fig:eval:geodomainStats}
\end{figure*}

Figure~\ref{fig:eval:geodomainTimeseries} shows the number of geo-domains queried for one of the traces. The number of geo-domains queried is consistently high throughout the trace with a median of 36. However, the number of uncached geo-domains that have to be queried from the DNS server is small with a median of 4. Most of the geo-domain queries are answered by the local cache as the device does not arbitrarily jump to radically different locations and query different geo-domains. 

Figure~\ref{fig:eval:geodomainsBox} shows the box plot of the total number of geo-domains queried across all records in all traces. Notice that most of the DNS requests are resolved from the cache. The long whiskers for both the cached and uncached box plots are because of the first \systemname{} query in the session. In the first query, all geo-domains need to be queried as the cache is empty. Figure~\ref{fig:eval:hitRatioDistribution} shows that the DNS cache hit ratio is close to 1 for a large number of \systemname{} queries.



\begin{figure}[!t]
    \centering
    \begin{subfigure}{.23\textwidth}
        \centering
        \includegraphics[width=.95\linewidth]{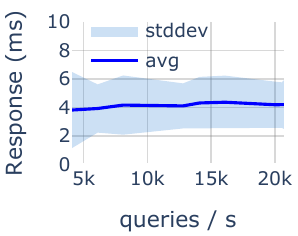}  
        \caption{Response latency.}
        \label{fig:eval:dnsServerLatency}
    \end{subfigure}
    \hfill
    \begin{subfigure}{.23\textwidth}
        \centering
        \includegraphics[width=.95\linewidth]{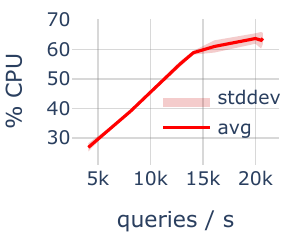}  
        \caption{CPU Usage.}
        \label{fig:eval:dnsServerCPU}
    \end{subfigure}
    \caption{DNS server capacity.}
    \label{fig:eval:dnsServerLoad}
\end{figure}

We used \texttt{dnsperf}~\cite{dnsperf} to benchmark our DNS server. The DNS requests were generated on a machine in the same local network as the DNS server. Figure~\ref{fig:eval:dnsServerLatency} shows that even at 20,000 DNS queries per second, the server response latency stays at 4~ms. This means that a single DNS server, without complex infrastructure set up, can support thousands of \systemname{} clients resolve geo-domain queries.

%% file: pages/evaluation/localization.tex
\subsection{Localization}
\label{subsec:eval:localization}

\begin{figure}[b]
    \begin{subfigure}{.23\textwidth}
        \centering
        \includegraphics[width=.95\linewidth]{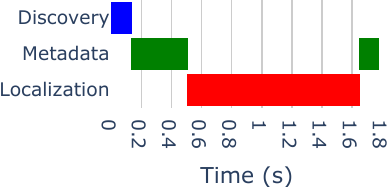}  
        \caption{Duration of network requests.}
        \label{fig:eval:requestsDurationZoom}
    \end{subfigure}
    \begin{subfigure}{.23\textwidth}
        \centering
        \includegraphics[width=.95\linewidth]{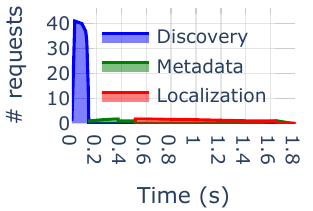}  
        \caption{Number of network requests.}
        \label{fig:eval:numRequestsZoom}
    \end{subfigure}
    \caption{Network activity of localization service.}
    \label{fig:discovery:networkActivity}
\end{figure}

The localization service needs to work in tandem with the discovery phase as the user moves through different maps. Figure~\ref{fig:discovery:networkActivity} shows the network activity on an \systemname{} client running the localization service. Figure~\ref{fig:eval:requestsDurationZoom} shows the time that the client spends in different phases in one cycle of queries including discovery and localization. Each horizontal bar represents the time from when the request was sent to when the response was received, including the time the client spent waiting for the response. Metadata refers to the negotiation of localization technologies. The client waits over 1 second for the localization result from the map servers. Note that other localization technologies could require far less time for the localization step of the process. In our AR application, while it waits for localization results, we continue to render AR content using the local tracking implemented by WebXR. We used the localization result to get an initial pose estimate and to correct drifts in local tracking. Therefore, the waiting time does not significantly affect the user experience. 

Figure~\ref{fig:eval:numRequestsZoom} show the number of requests sent in each phase. About 40 geo-domains are queried in the discovery phase as evidenced by the large number of network requests sent in the discovery phase. Despite a large number of requests, the time spent in the discovery phase is much shorter than in the localization phase (Figure~\ref{fig:eval:requestsDurationZoom}). This is because \systemname{} makes all DNS requests in parallel and the DNS server responds in a few tens of milliseconds. To capture the worst case for the number of requests sent in the discovery phase, we configure our DNS servers to prevent caching of NXDOMAIN results (i.e., negative results). However, as we show in \S~\ref{subsec:eval:discovery}, only the first request involves a large number of DNS queries, after which most geo-domains are cached.

\begin{figure}[t]
    \centering
    \includegraphics[width=.3\textwidth]{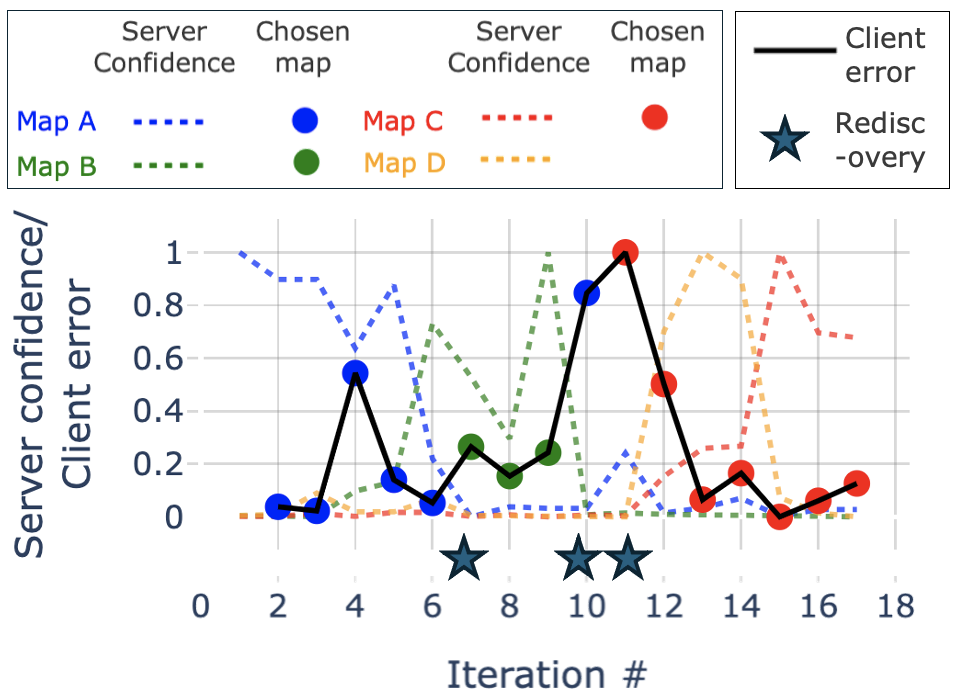}
    \caption{Map Selection.}
    \label{fig:eval:mapSelection}
\end{figure}

As described in \S~\ref{sec:localization}, \systemname{} does not run the discovery phase repeatedly but rather maintains an \texttt{activeServer} (Figure~\ref{fig:localization:flowchart}). To verify that \systemname{} returns the correct map server without having to run rediscovery every time, we set up 4 map servers, each covering a different indoor region in a university building. The regions were close enough that GPS location could not reliably distinguish between the 4 regions indoors. As a result, every time rediscovery is triggered, the addresses of all 4 map servers are returned and location cues are sent to all of them. The dotted lines in Figure~\ref{fig:eval:mapSelection} show the confidence of the map servers normalized between 0 and 1 as a reference. \systemname{} does not contact all servers every time, and therefore does not have access to this data but rather uses local VIO poses to estimate `client error'. The colored dots show the selected \texttt{activeServer}. We see that in most iterations, the \texttt{activeServer} is the server with the highest server confidence. The stars in the figure show that rediscovery roughly corresponds to the times when the confidence scores for the \texttt{activeServer} dips. Two consecutive rediscoveries are triggered when the confidence scores are low for all servers in a region where no map server has good coverage.

\input{pages/evaluation/localizationAccuracy}

%% file: pages/evaluation/routing.tex
\subsection{Routing}
\label{subsec:eval:routing}


We evaluate our implementation of the routing service (\S~\ref{subsec:routingService}) on an OpenStreetMap of a city. The routing service is run on two versions of the map -- a centralized version which retains the original form of the city map, and a sharded version where we split the map into a \textit{root city map} and smaller sub-maps. The smaller sub-maps consist of universities and some neighborhoods cut out of the city. We generated geo-domains for these sharded regions and registered them on our DNS server. These DNS records point to map servers providing routing service for each shard. We use the Open Source Routing Machine (OSRM)~\cite{luxen-vetter-2011} on the map servers to calculate optimal routes within maps. A routing query in the centralized version is made to the map server running OSRM on the whole city's map. In the sharded version, routing involves discovering relevant map servers using DNS queries and then requesting the map servers to run OSRM on their portion of the map.

\begin{figure}[b]
    \centering
    \begin{subfigure}{.23\textwidth}
        \centering
        \includegraphics[width=.95\linewidth]{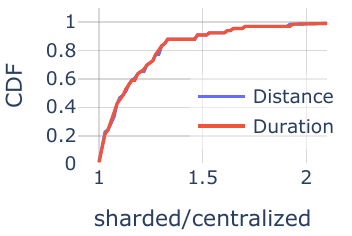}  
        \caption{Routing optimality ratios.}
        \label{fig:eval:routingOptimality}
    \end{subfigure}
    \hfill
    \begin{subfigure}{.23\textwidth}
        \centering
        \includegraphics[width=.95\linewidth]{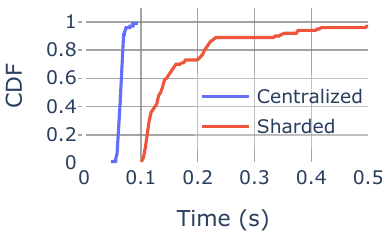}  
        \caption{Routing time taken.}
        \label{fig:eval:routingTime}
    \end{subfigure}
    \caption{Routing performance on Sharded maps.}
    \label{fig:eval:routingPerformance}
\end{figure}

Figure~\ref{fig:eval:routingOptimality} shows the ratio of the travel distance along the path returned by sharded and centralized routing. A ratio of 1 indicates that the distances are the same, while higher ratios show the extent of sub-optimality of sharded routing. The median ratio is 1.12 and the 90$\textsuperscript{th}$ percentile is 1.47. Figure~\ref{fig:eval:routingTime} shows the CDF of the times taken for the completion of routing queries on the sharded and centralized models. 
At the median, a sharded routing query takes about twice as long as a centralized query. 

%% file: pages/evaluation/reverseGeocode.tex
\subsection{Reverse Geocode}
\label{subsec:eval:reverseGeocode}

\begin{figure}
    \centering
    \includegraphics[width=.5\linewidth]{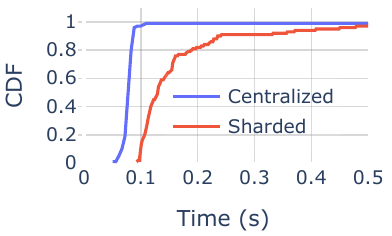}  
    \caption{Reverse Geocode performance.}
    \label{fig:eval:reverseGeocode}
\end{figure}

We evaluate our implementation of the reverse geocode service using the same setup of centralized and sharded maps as described in \S~\ref{subsec:eval:routing}. Figure~\ref{fig:eval:reverseGeocode} shows the CDFs of the time taken for reverse geocode queries on sharded and centralized maps. It includes both discovery and map data query time. At median, the ratio of sharded to centralized time is 1.67.

%% file: pages/relatedWork.tex
\section{Related work}

\paragraph{Federated mapping.} MapCruncher~\cite{elson2007mapcruncher} recognized the need to allow interoperability of distributed geographic data, focusing on layering interactive map data (called `mashups') from different sources. MapSynthesizer~\cite{elias2008live} built an application on top of MapCruncher that could discover and render tiles from distributed sources. They do not go into the details of how map data from different sources can be organized and discovered as their focus is on interactive map tiles. Our approach (introduced in~\cite{openflameHotos}) is to treat maps as an abstraction, and build an infrastructure to organize and discover them so they can expose any generic service (including visualization) to applications.

\paragraph{DNS-based discovery.} There have been some proposals in the past that have used the DNS for discovery of services. DNS-based Service Discovery (DNS-SD)~\cite{rfc6763} (eg. Apple Bonjour~\cite{appleBonjour}) uses the DNS to discover services of a given type under a given domain. In these systems, discovery is not based on geographic location but within a local network. DNS LOC records~\cite{rfc1876} store latitude and longitude as part of record data. However, these records only serve as metadata associated with a domain name and do not assist with location-based discovery. Several proposals have been made to extend the DNS to support geographic location-based queries, specifically for Vehicular Ad-hoc networks~(VANETs)~\cite{imielinski1999gps, fioreze2010extending, fioreze2011extending, moscoviter2016improving}. However, these proposals require altering DNS implementation and were not adopted. Recent work by Gibb et al~\cite{gibb2023earth} introduces location-aware DNS queries by leveraging the hierarchy in civic addresses of locations. It does not explore how to generate these domains to discover nearby services.



%% file: pages/conclusion.tex
\section{Conclusion}


In this paper we present the design and implementation of \systemname{}, a federated mapping system. \systemname{} organizes the world into smaller maps hosted on map servers maintained by disparate parties. It can incorporate private indoor maps, is scalable, and has a low barrier to entry. We implement a DNS-based map server discovery system so we can reuse existing DNS infrastructure and caching mechanisms. We also implement location-based services, including map visualization, geocoding, routing, and localization, on federated maps. We believe that a federated mapping system is essential for future spatial applications and hope that our paper acts as an impetus for the research community to start democratizing maps.

%% file: pages/appendix.tex
\vfill\break

\input{pages/design/designSpaceExploration}

\section{Application and Tools}
\label{sec:appendix_app}

\begin{figure} [h]
    \centering
    \includegraphics[width=0.3\textwidth]{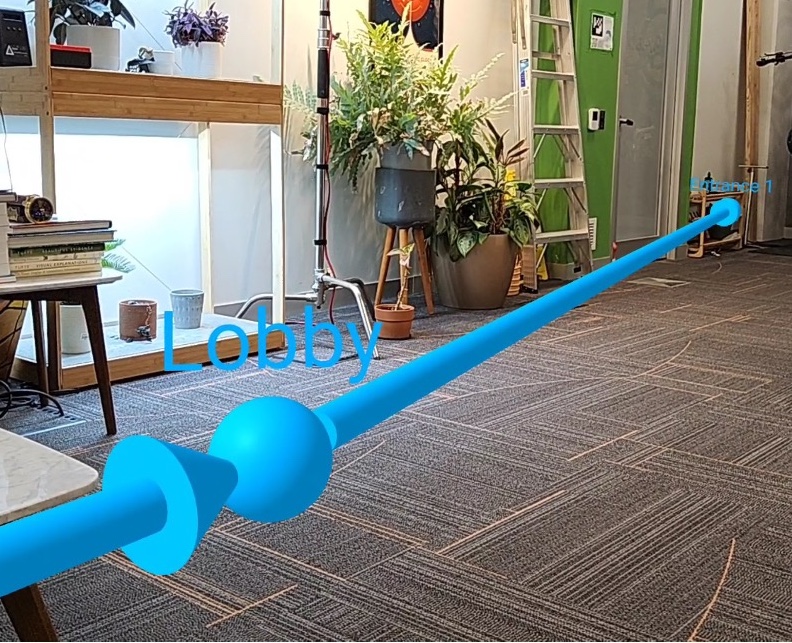}
    \caption{AR indoor navigation application built on \systemname{}.}
    \label{fig:application:navigationApp}
\end{figure}

Figure~\ref{fig:application:navigationApp} shows a screenshot of the 3D indoor AR navigation application built on \systemname{}. 

\begin{figure} [h]
    \centering
    \includegraphics[width=0.3\textwidth]{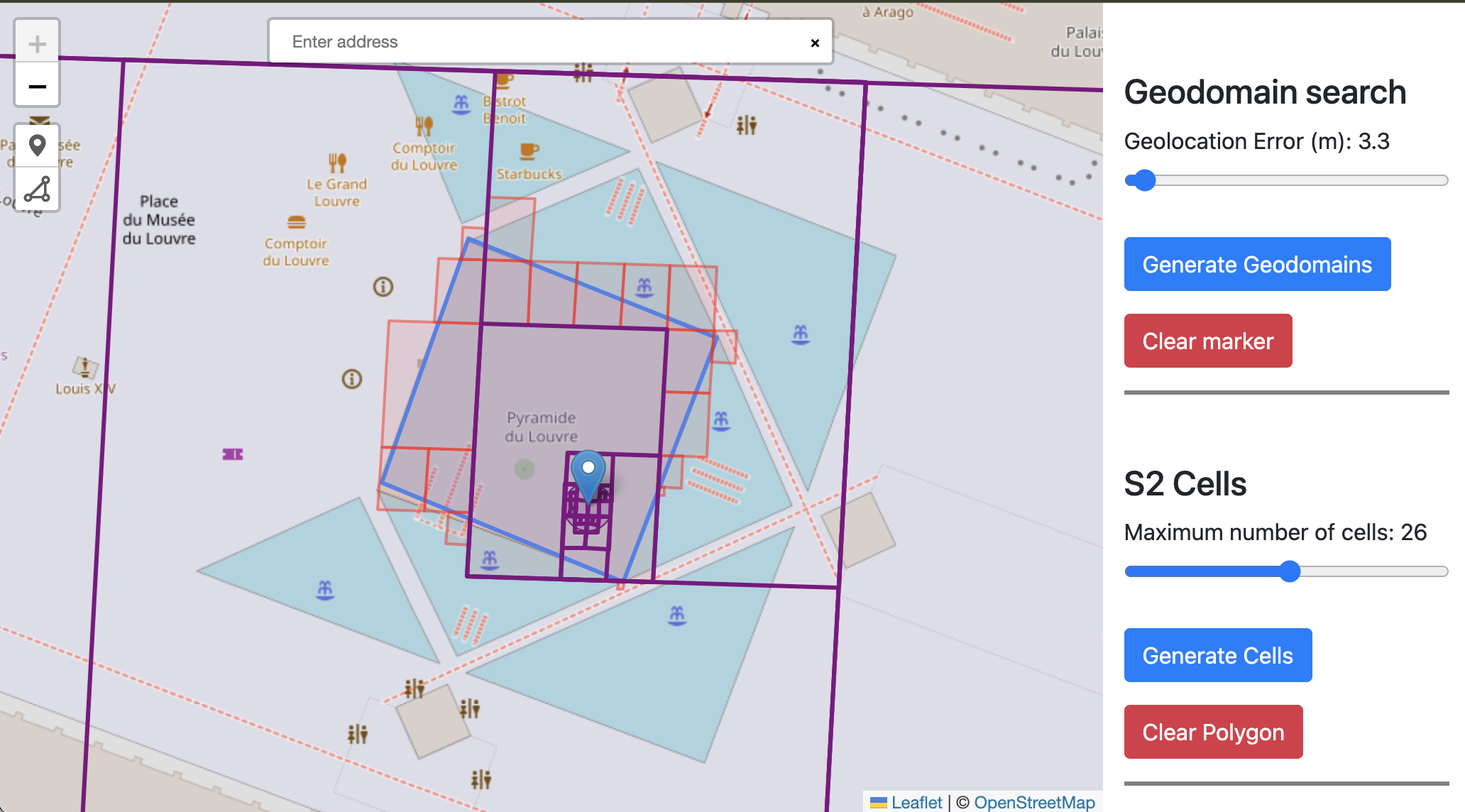}
    \caption{Geo-domain Explorer.}
    \label{fig:appendix:geodomainExplorer}
\end{figure}

The Geo-domain Explorer tool automates the process of generating DNS records for registering a map server or zone with the DNS. We have hosted it on \website{}. Figure~\ref{fig:appendix:geodomainExplorer} shows a screenshot of the tool.

\begin{figure} [h]
    \centering
    \includegraphics[width=0.3\textwidth]{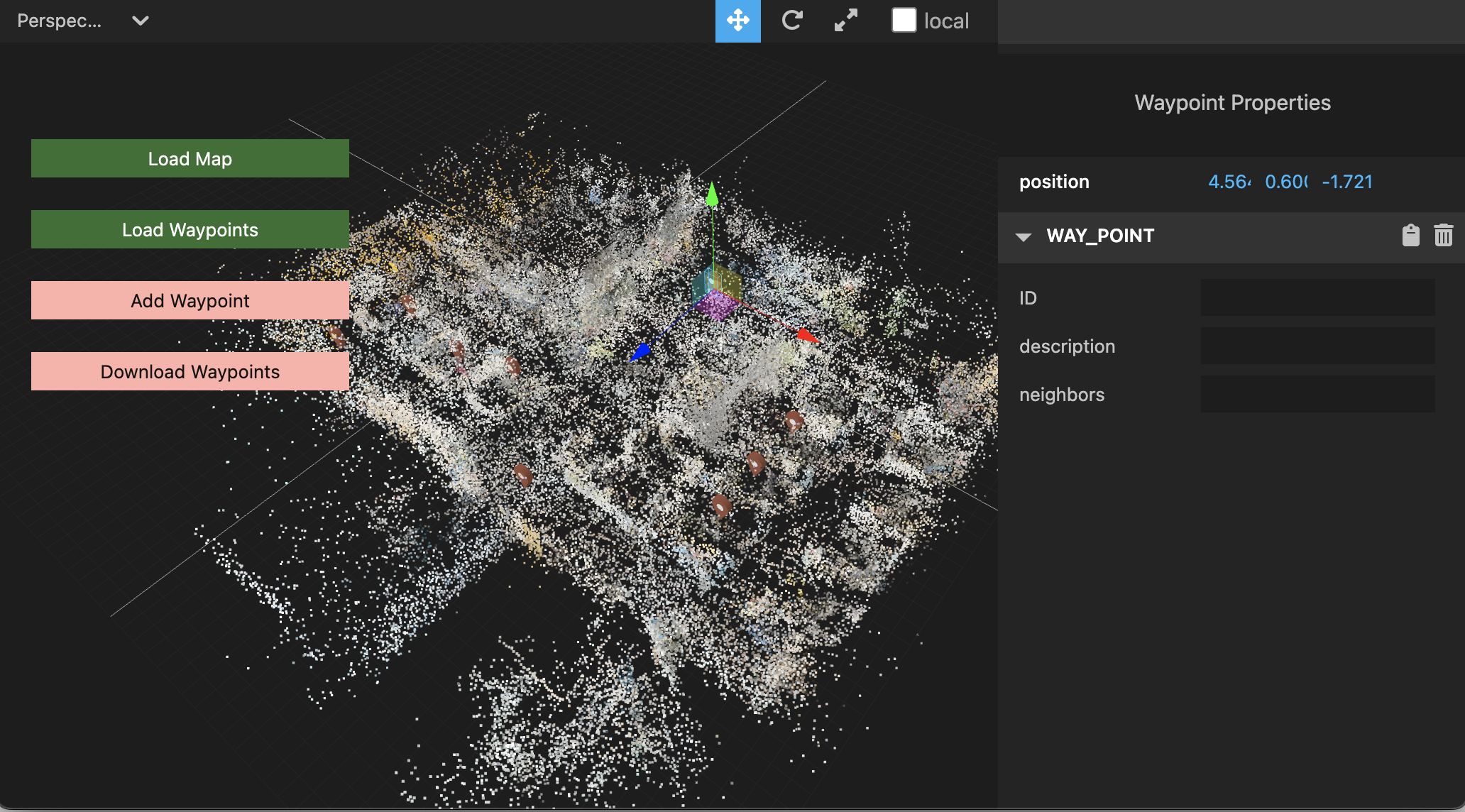}
    \caption{3D map creator.}
    \label{fig:appendix:waypointTagger}
\end{figure}

The Waypoint Tagger tool helps map creators tag map nodes and ways on their 3D scans. It also exports the map to be used with a map server. Figure~\ref{fig:appendix:waypointTagger} shows a screenshot from the tool.

\input{pages/locBasedServices/heterogenousMaps}

%% file: pages/design/designSpaceExploration.tex
\section{Design space exploration}
\label{subsec:designSpaceExploration}

A spatial naming system associates locations with \textit{names}. In the paper we choose \emph{maps} (i.e., labels associated with locations and relationships between labels) as names. However, there are alternate choices for names that can lead to a different system design.

\textit{Content-centric} design -- the naming system associates locations with content hosted at those locations. For example, at a supermarket, the naming system might discover the products sold at the supermarket. 
We believe that in future spatial applications, content will be dynamically generated at a high rate (as we see on Web applications today). Content-centric design will be bottle-necked at the naming system which is detrimental to spatial applications.
 
\textit{Application-centric} design -- the naming system associates locations with applications hosted at those locations. For example, at a supermarket, the naming system might discover the product search and navigation application. This is analogous to the Web where names (eg. \texttt{google.com}) refer to applications (eg. Web search). The applications maintain all of their content making the naming system independent of the amount of content generated, thereby removing a bottleneck from the system. As the content is spatial, applications need to maintain pointers maps that can provide context about the location of their content with respect to the real world. For example, to navigate users to their products of interest, the application needs a localization service that can determine the position of the user within the store. This inhibits incremental deployability of services. For example, consider the supermarket sharding its localization service to provide it separately for the clothing and electronics section, while decommissioning their store wide service. While this enhances ease of maintenance for the store, it breaks applications that still point to the older store wide service.

\textit{Map-centric} design -- The naming system associates locations with maps and the services provided on these maps. Related maps can be grouped into separate \textit{zones} and maintained autonomously, just as physical regions are maintained by disparate entities. This system is independent of content hosted against maps ensuring scalability. It also affords incremental deployability. Applications author their content with respect to map elements, use standardized interfaces to obtain map services and are agnostic to which machines are serving them. This is the design we choose for \systemname{}. Table~\ref{table:designMatrix} shows the matrix of system characteristics against design choices. 

\begin{table}
\resizebox{0.7\columnwidth}{!}{%
\begin{tabular}{l|ccc|}
\cline{2-4}
                                                                                             & \multicolumn{1}{l|}{\begin{tabular}[c]{@{}l@{}}Content\\ -centric\end{tabular}} & \multicolumn{1}{l|}{\begin{tabular}[c]{@{}l@{}}Application\\ -centric\end{tabular}} & \multicolumn{1}{l|}{\begin{tabular}[c]{@{}l@{}}Map\\ -centric\end{tabular}} \\ \hline
\multicolumn{1}{|l|}{Scalability}                                                            & \multicolumn{1}{c|}{{\color[HTML]{CB0000} \xmark}}                              & \multicolumn{1}{c|}{{\color[HTML]{036400} \cmark}}                                  & {\color[HTML]{036400} \cmark}                                               \\ \hline
\multicolumn{1}{|l|}{\begin{tabular}[c]{@{}l@{}}Incremental\\ deployability\end{tabular}}    & \multicolumn{1}{c|}{{\color[HTML]{CB0000} \xmark}}                              & \multicolumn{1}{c|}{{\color[HTML]{CB0000} \xmark}}                                  & {\color[HTML]{036400} \cmark}                                               \\ \hline
\multicolumn{1}{|l|}{\begin{tabular}[c]{@{}l@{}}Space ownership\\ Enforcement\end{tabular}}  & \multicolumn{3}{c|}{}                                                                                                                                                                                                                               \\ \cline{1-1}
\multicolumn{1}{|l|}{Delegability}                                                           & \multicolumn{3}{c|}{\multirow{-2}{*}{\begin{tabular}[c]{@{}c@{}}Depends on implementation \\ of the discovery system\end{tabular}}}                                                                                                                 \\ \hline
\multicolumn{1}{|l|}{\begin{tabular}[c]{@{}l@{}}Low barrier for\\ map creators\end{tabular}} & \multicolumn{3}{c|}{}                                                                                                                                                                                                                               \\ \cline{1-1}
\multicolumn{1}{|l|}{\begin{tabular}[c]{@{}l@{}}Low expectation\\ from clients\end{tabular}} & \multicolumn{3}{c|}{\multirow{-2}{*}{\begin{tabular}[c]{@{}c@{}}Depends on implementation \\ of location-based services\end{tabular}}}                                                                                                              \\ \hline
\end{tabular}%
}
\caption{Design choices vs. system characteristics.}
\label{table:designMatrix}
\end{table}

%% file: pages/locBasedServices/heterogenousMaps.tex
\section{Map Services on Heterogeneous Maps}
\label{appendix:heterogenousMaps}

Map services need to support a heterogeneous set of maps using different coordinate systems. We call maps laid out in the global geographic system \textit{geo-based} maps. \textit{Local} maps (maps in their own local coordinate system) can be further classified based on how they label areas shared with surrounding maps. If a map uses the same labels for shared areas as the other maps around them, they can be explicitly stitched together, so we call them \textit{bound} maps. Otherwise we call them \textit{unbound}. The discovery system is agnostic to this heterogeneity and discovers all maps and services for a given location.

\begin{table}
\resizebox{\columnwidth}{!}{%
\begin{tabular}{l|c|l|l|}
\cline{2-4}
                                      & \multicolumn{1}{l|}{Geo-based} & Bound                         & Unbound                       \\ \hline
\multicolumn{1}{|l|}{Geocode}         & {\color[HTML]{036400} \cmark}  & {\color[HTML]{036400} \cmark} & {\color[HTML]{036400} \cmark} \\ \hline
\multicolumn{1}{|l|}{Reverse-geocode} & {\color[HTML]{036400} \cmark}  & {\color[HTML]{680100} \xmark} & {\color[HTML]{680100} \xmark} \\ \hline
\multicolumn{1}{|l|}{Routing}         & {\color[HTML]{036400} \cmark}  & {\color[HTML]{036400} \cmark} & {\color[HTML]{CD9934} \cmark} \\ \hline
\multicolumn{1}{|l|}{Localization}    & {\color[HTML]{036400} \cmark}  & {\color[HTML]{036400} \cmark} & {\color[HTML]{036400} \cmark} \\ \hline
\multicolumn{1}{|l|}{Tile rendering}  & {\color[HTML]{036400} \cmark}  & {\color[HTML]{036400} \cmark} & {\color[HTML]{CD9934} \cmark} \\ \hline
\end{tabular}%
}
\caption{Map services that can be provided on different kinds of maps. {\color[HTML]{036400} \cmark} shows the service can be provided on the map and its results can be combined with other maps if needed. {\color[HTML]{CD9934} \cmark} shows the service can be provided on the map but its results cannot be combined with other maps. {\color[HTML]{680100} \xmark} means the service cannot be provided on the map.}
\label{table:servicesOnMaps}
\end{table}

Table~\ref{table:servicesOnMaps} shows the different services that can be supported by each map type. Geo-based maps can support all services. Location-to-address conversion (reverse-geocoding) cannot be supported on local maps as they do not have a notion of latitudes and longitudes. Unbound maps can support routing within the map, but the routes calculated in such maps cannot be combined with paths from other maps as there are no common nodes.